\documentclass[useAMS,usenatbib,referee]{mn2e}
\usepackage{graphicx}
\usepackage[intlimits]{amsmath}
\usepackage{amssymb}

\begin{document}

\title[{\it INTEGRAL} SS433]
{{\it INTEGRAL} observations of SS433: system's parameters and nutation of supercritical
accretion disk}

\author[A.M.Cherepashchuk et al.]{A.M. Cherepashchuk$^{1}$, R.A. Sunyaev$^{2}$, 
S.V. Molkov$^{2}$, 
E.A. Antokhina$^{1}$, 
\and
K.A. Postnov$^{1}$, 
A.I. Bogomazov$^{1}$ 
\thanks{E-mail: cherepashchuk@gmail.com(AMCh);
kpostnov@gmail.com(KAP);elant@sai.msu.ru(EAA)} \\
$^{1}${\sl Moscow M.V. Lomonosov State University, Sternberg Astronomical Institute, Universitetskij pr. 13, Moscow 119992, Russia}\\ 
$^{2}${\sl Space Research Institute, Moscow, Russia}\\
}
\date{Accepted ......  Received ......; in original form ......
}

\maketitle

\begin{abstract}
Based on multiyear {\it INTEGRAL} observations of SS433 in 2003-2011, a composite IBIS/ISGRI 18-60 keV orbital 
light curve is constructed around zero precessional phases $\psi_{pr}= 0$ at the maximim
accretion disk opening angle. 
It shows a peculiar shape with
significant excess near the orbital phase $\phi_{orb}= 0.25$, which is not seen in the softer 2-10 keV energy band. 
The 40-60 keV orbital light curve demonstrates two almost equal 
humps at phases $\sim 0.25$ and
$\sim 0.75$, most likely due to nutation effects of the accretion disk. 
The nutational variability of SS433 in 15-50 keV with a period of $\simeq 6^d.290$ 
is independently found from analysis of \textit{Swift/BAT} data. 
The
change of the off-eclipse 18-60 keV X-ray flux with the precessional phase 
shows a double-wave form with strong primary maximum at $\psi_{pr}= 0$ and weak but significant
secondary maximum at $\psi_{pr}= 0.6$. A weak variability of the 18-60 keV flux in the middle of the orbital eclipse correlated with the disk precessional phase is also
observed. 
The joint analysis of the broadband 18-60 keV 
orbital and precessional light curves  
confirms the presence of a hot extended corona in the central parts of the
supercritical accretion disk and constrains the 
binary mass ratio in SS433 in the range $0.5\gtrsim q\gtrsim 0.3$, 
suggesting the black hole nature of the compact object. 
\end{abstract}


\section{Introduction}
\label{sec:intro}

SS433 is a unique galactic 
steadily superaccreting microquasar with mildly relativistic ($v=0.26c$), 
precessing jets located at a distance of 5.5 kpc \citep{Margon84, Cher81, Cher88, Fab04}.
The system exhibits three photometric and
spectral periodicities related to precession
($P_\mathrm{pr}\simeq162^d.5$), orbital ($P_\mathrm{orb}\simeq13^d.082$) and
nutation ($P_\mathrm{nut}\simeq6^d.29$) periods \citep{Goransk98}, which 
are found to be stable during an observational period of more than 30 years
\citep{Davydov08}.  
Despite the wealth of observations, the nature of the compact star in 
SS433 remains inconclusive. The presence
of absorption lines in the optical spectrum of the companion \citep{Gies02,hillwig08} suggests its spectral classification as $\sim$ A7Ib supergiant. 
Assuming these lines to be produced in the optical
star photosphere, their observed orbital Doppler shifts 
would correspond to the mass ratio of
compact ($M_x$) and optical ($M_v$) star $q=M_x/M_v\sim 0.3\pm
0.11$ and masses $M_x=(4.3\pm 0.8) M_\odot$, $M_v=(12.3\pm3.3) M_\odot$, respectively,
pointing to the black hole nature of the compact star. 

Modeling of all {\it INTEGRAL} eclipses of the source available before 2010 
using a purely geometrical model 
yeilded independent constraints on   
the binary mass ratio $q=0.25-0.5$ with the most probable value $q=0.3$, suggesting
the mass of the compact companion $M_x\simeq 5.3 M_\odot$ and the optical star 
$M_v\simeq 17.7 M_\odot$ for the observed optical star mass function 
$f_v=0.268 M_\odot$ \citep{Cher09, Cher13}. 
This firmly places SS433 among black-hole high-mass X-ray binaries (HXMB). 
Thus SS433 can be the only known example of galactic HMXB at an advanced
evolutionary stage \citep{Cher81, Cher88} with supercritical
accretion \citep{ShS73} onto a black hole.
Its study in different spectral bands provides invaluable information for 
theory of evolution of binary star and the formation of relativistic jets.  
However, adopting the spectral classification A7Ib does not 
rule out the presence of neutron star in SS433 \citep{Goransky_2011}. 
Preliminary analysis of 
the new optical spectroscopic observations by SS433 on large telescopes 
may raise doubts on the correctness of the 
spectral classification of the optical star, which apparently relates to a very powerful
mass loss from the optical star and the accretion disk in this system. Clearly, new observations of SS433 and independent determination of the binary system parameters are needed.

The basic picture of hard X-ray emitting regions, 
as emerged from analysis of X-ray data \citep{Ant92, Fab04, Fil06, Cher03, Cher05, Cher09, Krivosheev09}, includes 
hot X-ray jet propagating through a funnel in the supercritical 
accretion disk, filled with hot scattering medium (a corona).  
The X-ray spectrum of SS433 in the 3-100 keV range 
can be fitted by two-component model (thermal X-ray emission from 
the jet and thermal comptonization spectrum from corona) 
elaborated in \citep{Krivosheev09}.  The scattering corona 
parameters are: $T_{cor}\simeq 20$~keV, Thomson optical depth $\tau_T\simeq 0.2$
and mass outflow rate in the jet $\dot M_j=3\times 10^{19}$~g/s.
This parameters suggest the coronal electron number density around $5\times 10^{12}$ cm$^{-3}$, which is
typical in the wind outflowing with a velocity of $v\sim 3000$ km/s from a supercritical 
accretion disk with mass accretion rate onto the compact star $\dot M\sim 10^{-4}$
M$_\odot$/yr at distances $\sim 10^{12}$ cm from 
the center, where a Compton-thick photosphere is formed \citep{Fab04}.
The size of the disk photosphere was independently estimated
from measurements of fast optical aperiodic variability \citep{Burenin2010}. 

Here we analyze hard X-ray eclipses of SS433 near moment T3, corresponding to 
maximum separation of the moving emission lines from jets and the maximum opening angle
of the accretion disk, in 
combination with the precessional variability
as observed by {\it INTEGRAL} in 2003-2011, and treat them in terms of our multicomponent 
geometrical model \citep{Cher09}. In addition to
precessional and orbital hard X-ray variability, which allowed us to constrain
the parameters of the system within the adopted geometrical model \citep{Cher09}, here we focus on the features in the observed hard X-ray light curve 
which can be interpreted in terms of nutation of the supercritical accretion disk
in SS433 \citep{Cher13}. 

In Section \ref{s_observ} we describe 
{\it INTEGRAL} observations used for the analysis. 
In Section \ref{s_precvar} 
we discuss the precessional variability of SS433.
In Section \ref{s_primecl} we study the primary 
eclipse of SS433 in hard X-rays and present IBIS/ISGRI spectra
of SS433 obtained at different orbital phases. 
In Section \ref{s_nut} we consider evidence for nutational 
variability of SS433 from \textit{Swift/BAT} and {\it INTEGRAL} observations. 
We discuss the results in Section \ref{s_disc} and summarize
our findings in Section \ref{s_concl}.

\section{Observations}
\label{s_observ}

Dedicated {\it INTEGRAL} observations of hard X-ray eclipse are summarized in
Table \ref{t:obs}. \textbf{The observations have been carried out using $5\times5$ 
dithering mode, so only IBIS/ISGRI data havee been analyzed}. 
The observations were mostly concentrated around precessional 
phase zero ('the T3 moment' in terms of the kinematic model of SS433 \citep{Cher81}), where the accretion disk is maximum opened to the observer
and the non-eclipsed X-ray flux from the source is the highest, reaching typically 
$\sim 20$~mCrab in the 20-60 keV band. 

\begin{table*}
\caption{Dedicated {\it INTEGRAL} observations of SS433 primary 
eclipses at precessional phases with maximum accretion disk opening \label{t:obs}}
\footnotesize
\centering
\begin{tabular}{lllc}
\hline
Set&{\it INTEGRAL} orbits& Dates & Precessional phase $\psi_{pr}$\\
\hline
I&67-70 & May 2003 &0.001-0.060\\
II&555-556 & May 2007 &0.980-0.014\\
III&608-609 &October 2007 &0.956-0.990\\
IV&612-613 &October 2007 &0.030-0.064\\ 
V&722-723 &September 2008&0.057-0.091\\
VI&984     &November 2010&0.870-0.890\\ 
&987     &November 2010&0.930-0.940\\
VII& 1040-1041&April 2011&0.910-0.950\\
\hline 
\end{tabular}
\end{table*}

\begin{figure*}
\includegraphics[width=0.9\textwidth]{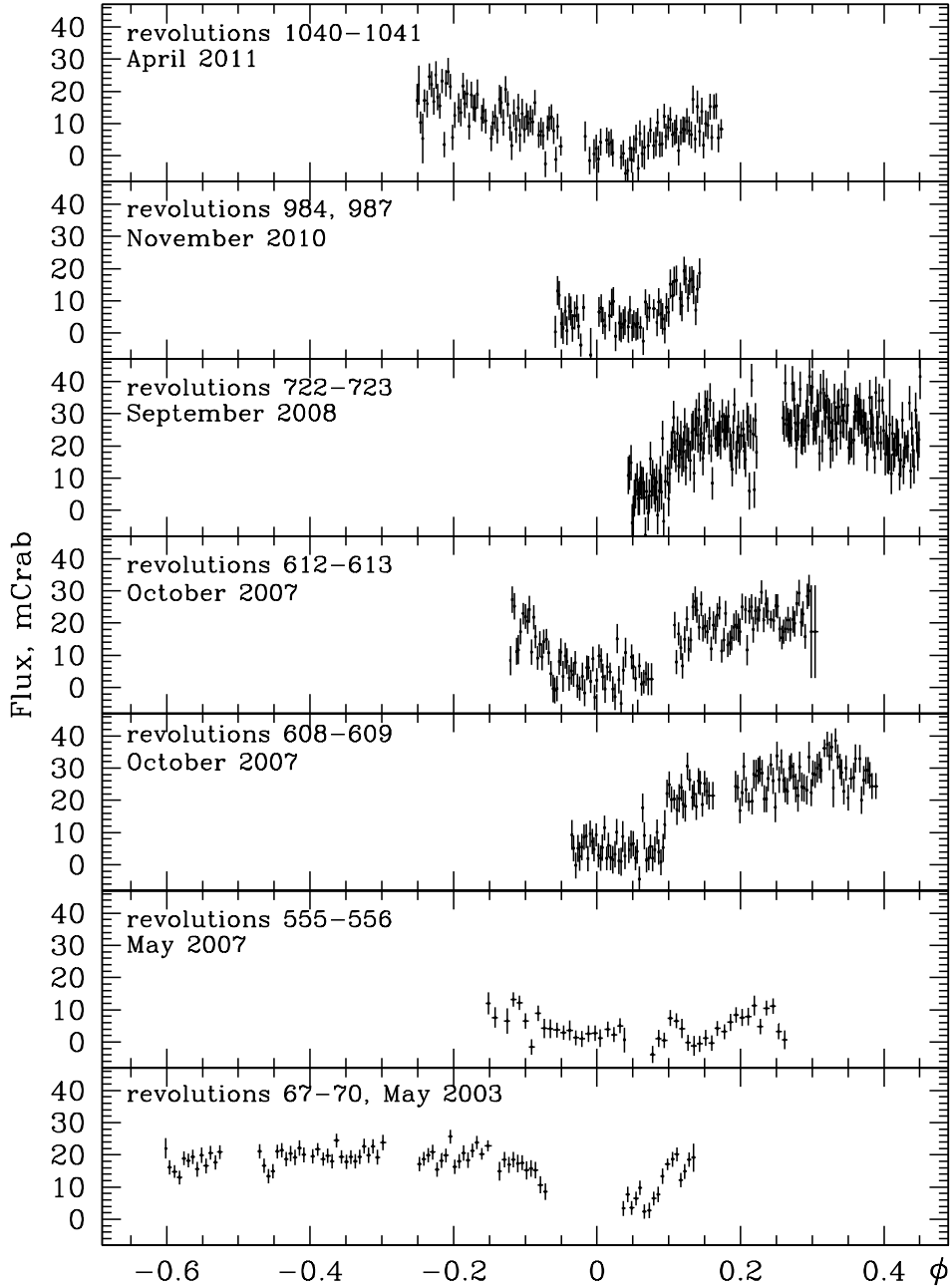}
{\caption{Primary eclipses of SS433 near precession phase zero observed by
{\it INTEGRAL} (observations I-VII from Table 1). 
18-60 keV IBIS/ISGRI fluxes obtained in  
individual science windows are shown.}\label{f:alldata}}
\end{figure*}

\textbf{The {\it INTEGRAL} data were processed using the original software
package elaborated by the IKI {\it INTEGRAL} team for the IBIS/ISGRI telescope image
reconstruction (see \citep{Mol04, Revnivtsev_al04, Krivonos_al10} for more detail).}
The 18-60 keV light curves  of SS433 eclipses 
obtained in observations I-VII near 
precession phase  zero (moment T3),  
from Table 1 are presented in Fig. \ref{f:alldata}.

\section{Precessional variability}
\label{s_precvar}

We start with producing the observed precessional light curve of SS433
from all available {\it INTEGRAL} data taken outside the primary eclipses.
For the analysis of precessional variability of SS433 
we have used both data from our {\it INTEGRAL} observing program of
SS433 and publically available data of all {\it INTEGRAL} observations where the
source fell in the field of view of the IBIS/ISGRI telescope ($<13$ degrees). The
total exposure time of the selected data is approximately $8.5$ Ms. 
To perform
precessional-phase-resolved analysis we ascribed to each SCW (Science Window
or SCW, the natural piece of the {\it INTEGRAL} data -- pointing observation with
an exposure time of $\sim2-5$~ks) the appropriate orbital and precessional phases.
The phases have been calculated using the ephemeris from \citep{Fab04}: 

the orbital primary minimum (corresponding to $\phi_{orb}=0$, when the optical star is in 
front of the disk) 
$$
JD_{MinI} \hbox{(hel)} = 2450023.62 + 13.08211\times E\,,
$$

the zero precession phase (corresponding to the T3  moment, $\psi_{pr}=0$, when 
the disk is maximum open to the observer)
$$
JD_{T3} \hbox{(hel)}= 2443507.47 + 162.375\times E1\,.
$$

\begin{figure*}
\includegraphics[width=0.7\textwidth]{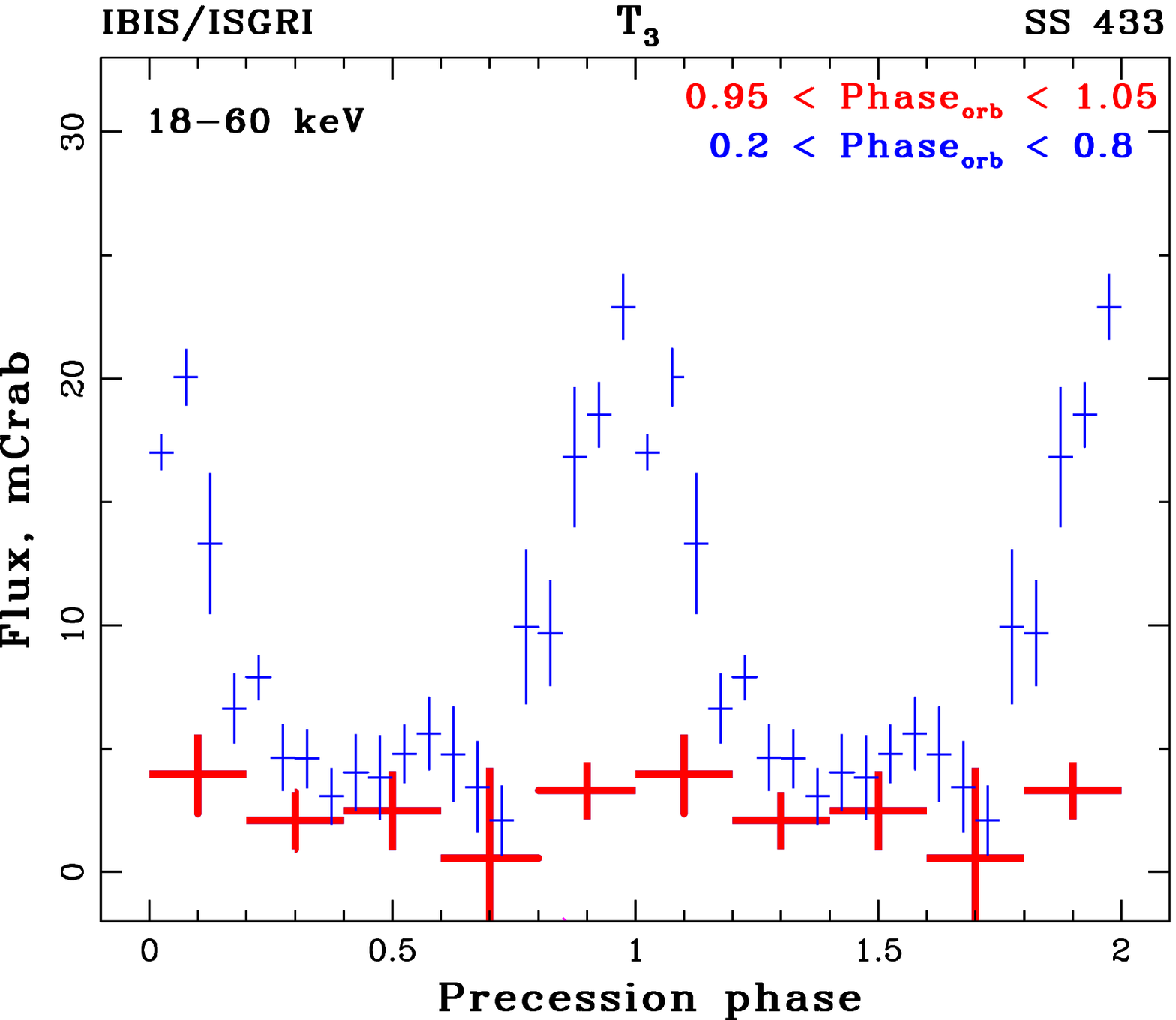}
{\caption{Precessional 18-60 keV light curves out of the primary orbital eclipses (upper thin crosses)
and inside the eclipses (thick gray crosses)}\label{f:hprecmin}}
\vfill
\includegraphics[width=0.7\textwidth]{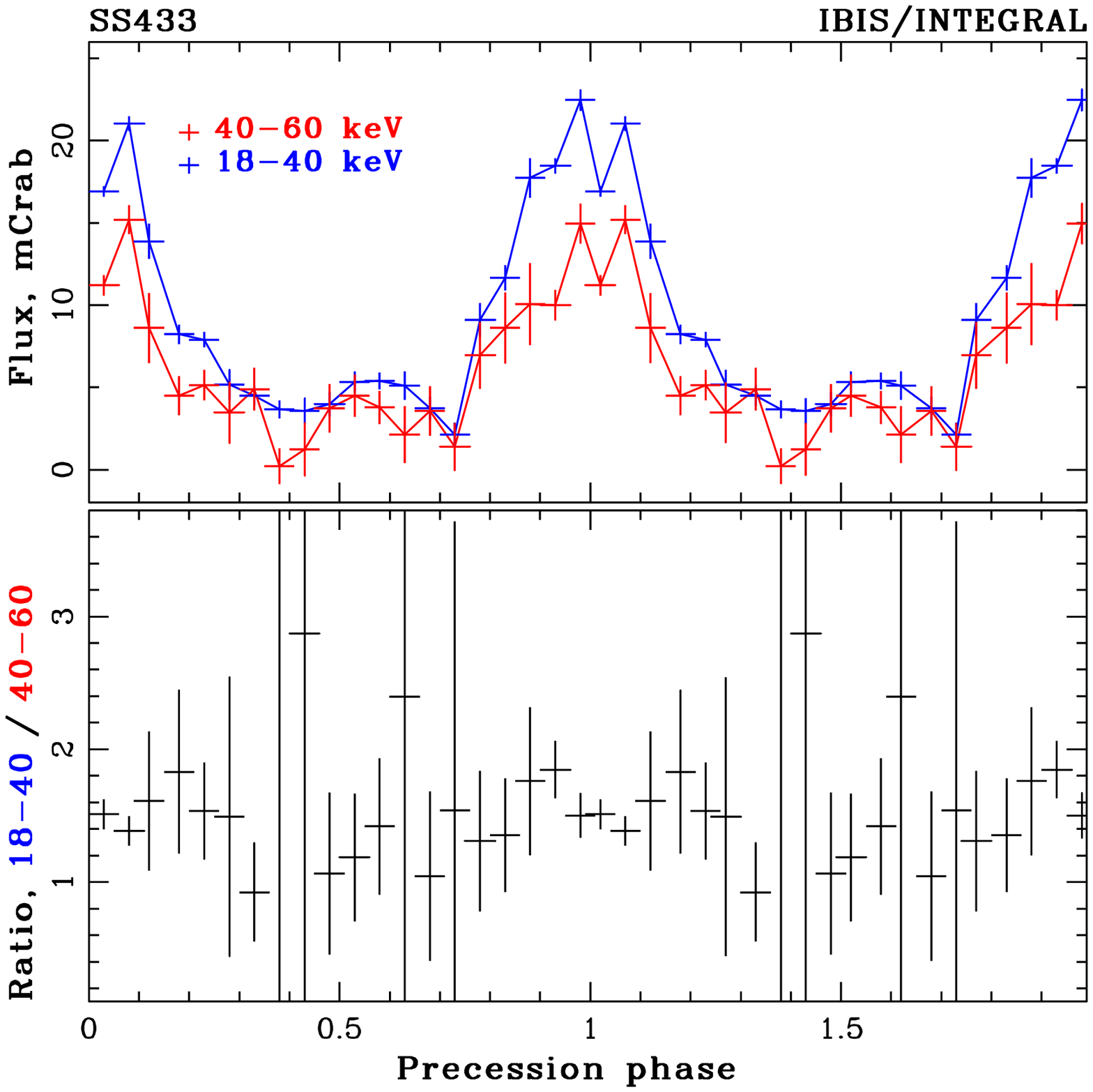}
{\caption{Precessional 18-40/40-60 keV light curves (upper panel) and the hardness ratio (bottom panel).
Orbital eclipses are excluded.}\label{f:hprec}}
\end{figure*}


At a distance of 5.5 kpc to SS433, as derived from the kinematic
properties of moving emission lines \citep{Fab04}, the observed 18-60 keV X-ray flux  
corresponds to a maximum uneclipsed hard X-ray luminosity of 
$3\times 10^{35}$~erg/s. The precessional change of the X-ray flux is shown in 
Fig. \ref{f:hprec}.  To plot this Figure, all available observations of
SS433 by {\it INTEGRAL} with a total exposure time of about 8.5 Ms were used.  
The precession light curves were separately constructed for uneclipsed ($0.2<\phi_{orb}<0.8$)
and middle-eclipse ($0.95<\phi_{orb}<1.05$) observations.
Light curves in both bands 
show precessional variability and were used to constrain binary system 
parameters (see \citet{Cher09} and Section \ref{s:param} below). 

The hard X-ray precessional variability of the source has been 
found to be quite significant and stable over several 
precessional periods. The maximum to minimum flux ratio of the average precessional variability 
is around 5-7, which is higher than that in softer X-ray bands, and suggests 
the hard X-ray emission originating closer to the base of the visible part of the jets.
The secondary maximum at the precessional phase $\sim 0.6$ is clearly seen in
both 18-40 keV and 40-60 keV light curves (Fig. \ref{f:hprec}). The precessional
18-60 keV light curve constructed for data taken at the middle of X-ray eclipse 
(Fig. \ref{f:hprecmin}) has a non-zero flux at all precessional phases, 
as expected from a hot scattering corona above the accretion disk.  
  
\begin{figure*}
\includegraphics[width=0.47\textwidth]{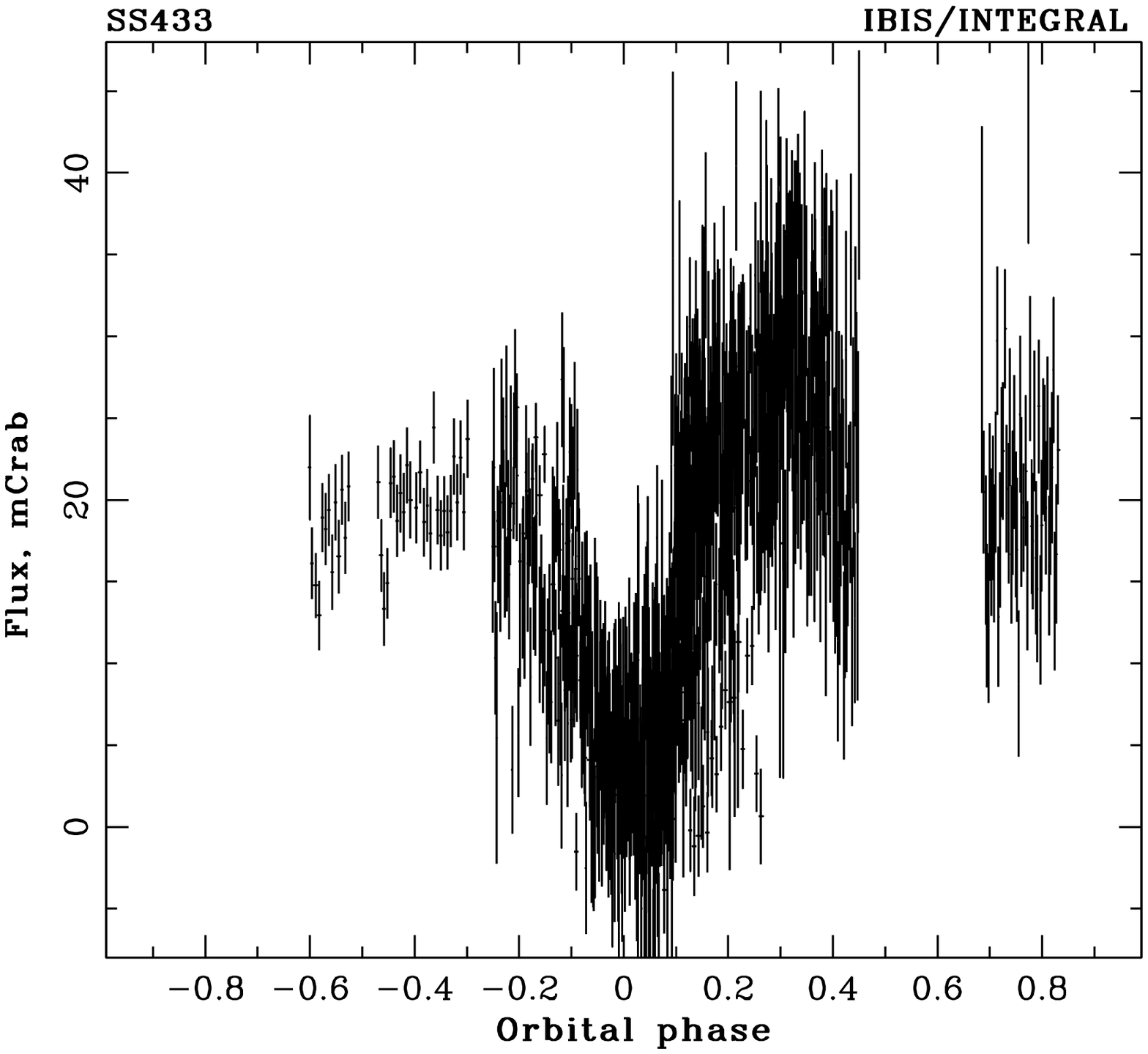}
\hfill
\includegraphics[width=0.47\textwidth]{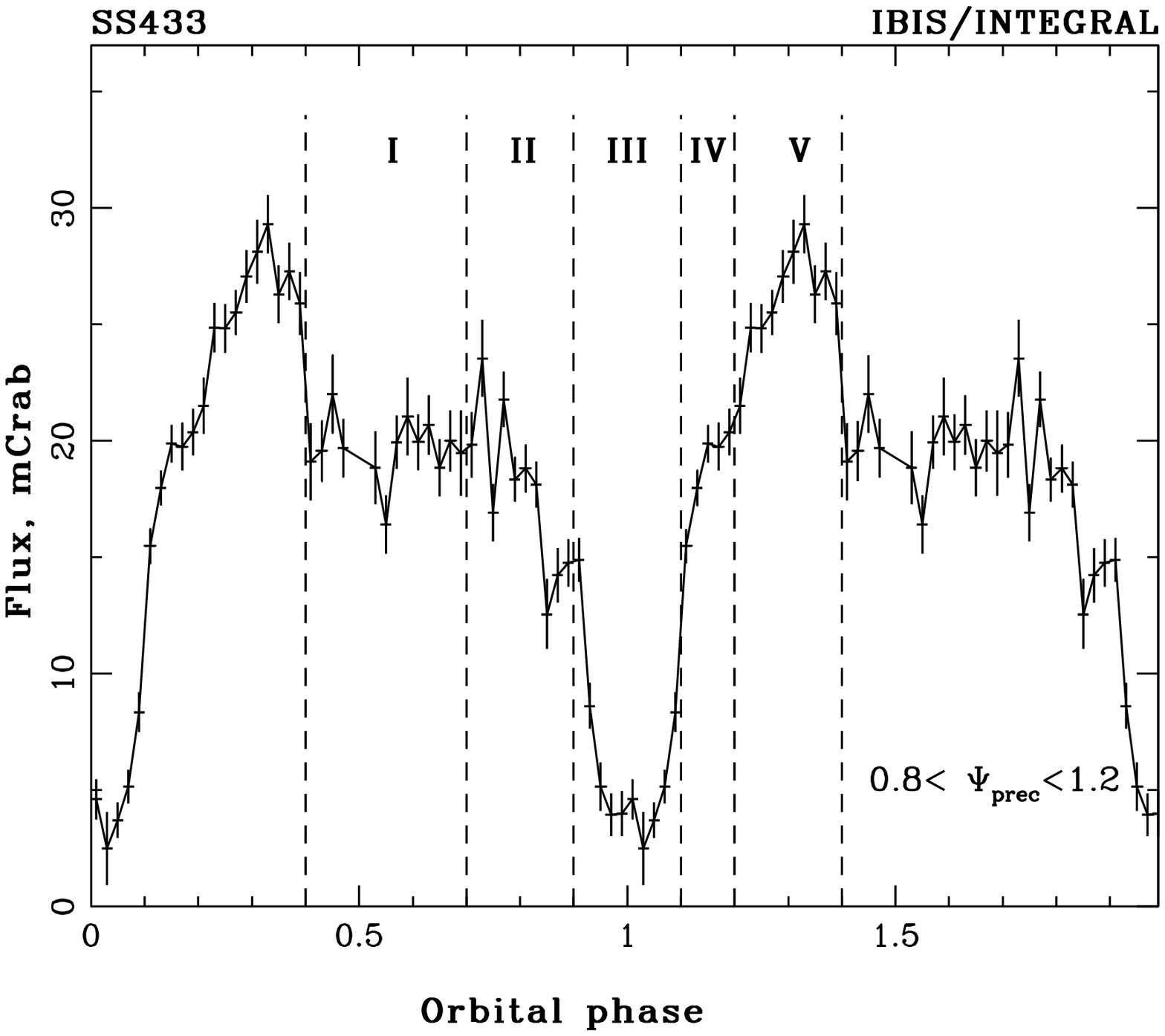}
\parbox[t]{0.47\textwidth}{\caption{Composite IBIS/ISGRI 18-60 keV X-ray eclipse 
light curve around precessional phase zero ($0.08 < \psi_{pr} <1.2$) for observations in Table 1.}\label{f:lc}}
\hfill
\parbox[t]{0.47\textwidth} 
{\caption{Binned X-ray light curve ($\Delta \phi_{orb}=0.02$) with orbital  
phase intervals I-V for spectral analysis (vertical dashed lines). 
}\label{f:lcbinned}}
\end{figure*}

\subsection{Orbital variability}
\label{s_primecl}

\begin{figure*}
\includegraphics[width=0.6\textwidth]{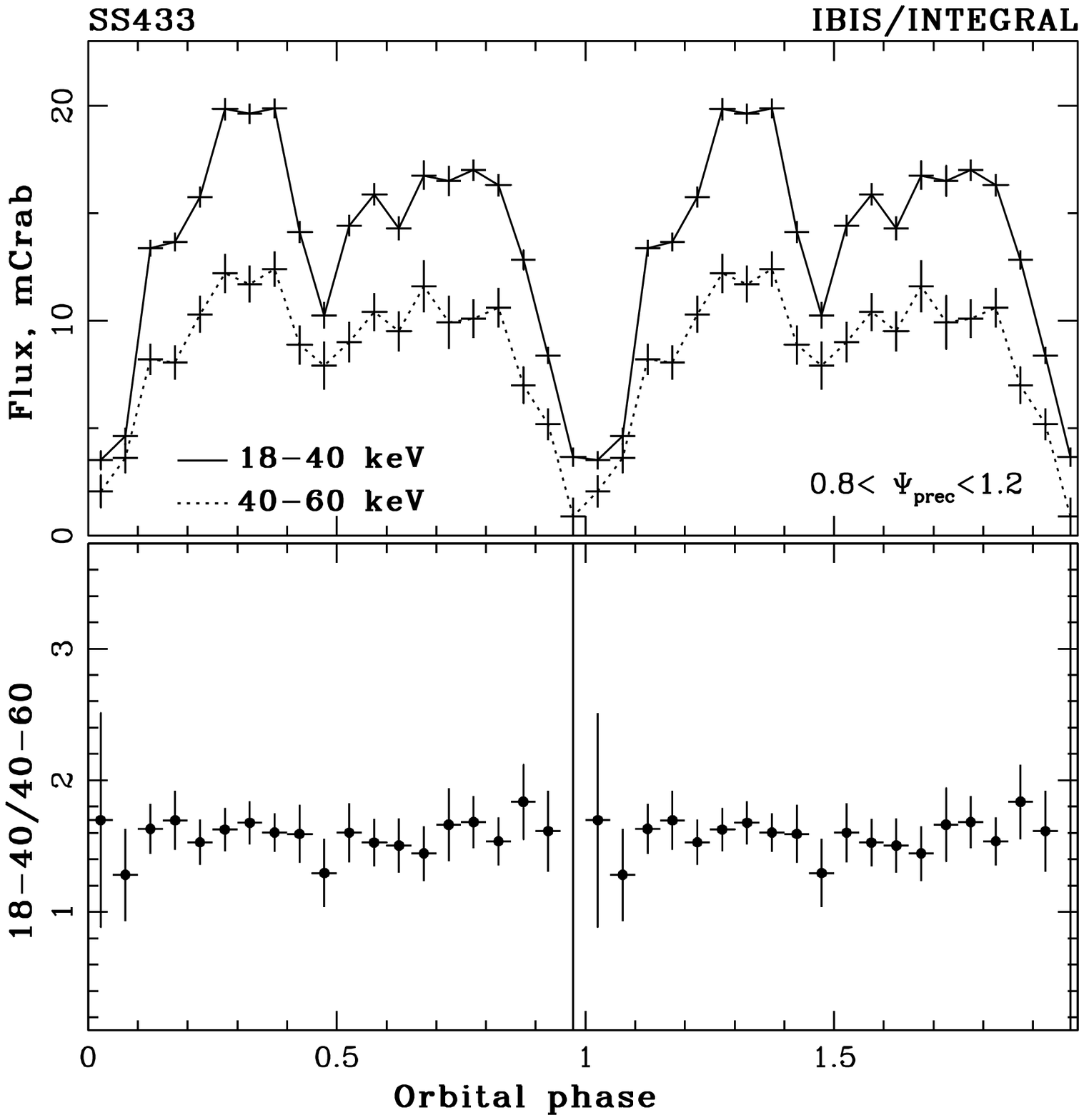}
{\caption{IBIS/ISGRI 18-40 and 40-60 keV orbital light curves (the upper and
lower curves, respectively, on the upper panel) with 
the hardness
ratio (bottom panel).}\label{f:hrorb}}
\vfill
\includegraphics[width=0.6\textwidth]{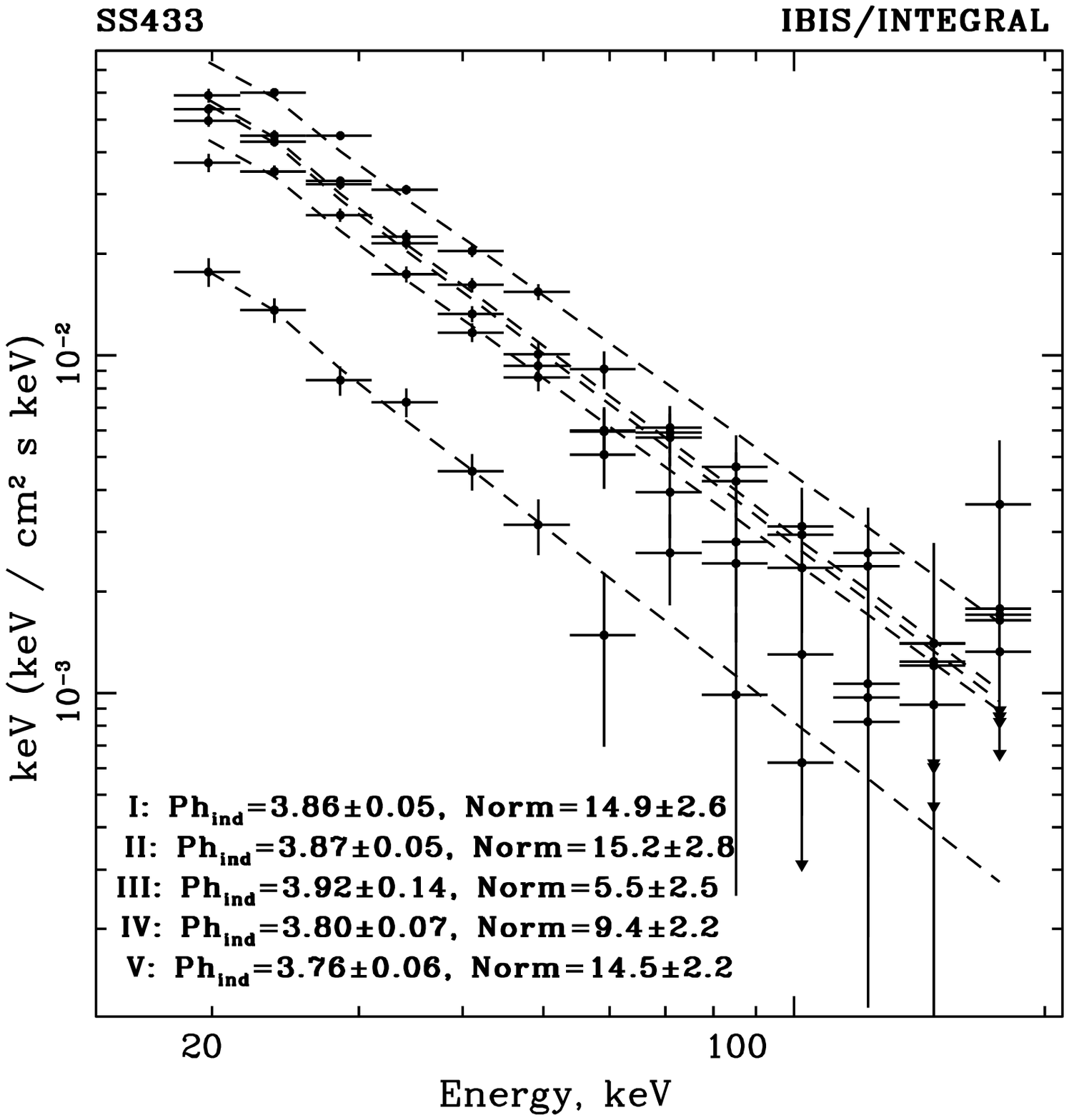}
{\caption{ Phase-resolved IBIS/ISGRI 
spectra of SS433 within chosen orbital phase intervals I-V shown in Fig. \ref{f:lcbinned}.
1-$\sigma$ errors are shown.}\label{f:spectra}}
\end{figure*}

The composite IBIS/ISGRI 18-60 keV light curve of the primary X-ray eclipse
at precessional phase zero (sets I-VII from Table \ref{t:obs}) 
is shown in Fig. \ref{f:lc}. Data were analyzed 
using the IKI {\it INTEGRAL} data processing code described in \citet{Mol04}. 
\textbf{The 
variable character of emission from SS433 is clearly seen in Fig. 4. For example, some eclipses (e.g. in May 2007, shown in Fig. 1), are atypically broad, and we have not included them in our analysis. Fig. 4 clearly demonstrates that the eclipse ingress is more stable than the eclipse egress, and we specially stressed this point in previous publications \citep{Cher09}. The variability of eclipse egresses, likely reflecting the complex geometry of gas flows in SS433, makes the light curve modeling very difficult. However, the flux averaging within orbital phase bins $\Delta \phi_{orb}=0.02$ increases the signal-to-noise ratio.}  
The
eclipse light curve averaged within orbital phase bins $\Delta \phi_{orb}=0.02$ is 
shown in Fig. \ref{f:lcbinned}. 3$\sigma$-flux errors are indicated.
The 18-40 and 40-60 keV orbital
light curves with the corresponding hardness ratios are 
presented in Fig. \ref{f:hrorb}.  
\textbf{Figs. 5 and 6 show that the averaged 
eclipse light curve of SS433 demonstrates 
regular features at orbital phases $\sim 0.2$ and $\sim 0.7$,  
which we will discuss in the next Section}.

Earlier we have shown \citep{Cher09} that the form of hard X-ray spectrum 
of SS433 does not change with the precessional phase: at maximum and minumum of 
the precessional variability, the spectrum is fitted by a power-law 
$dN/dE/dt/dA\sim E^{-\Gamma_{ph}}$ 
with the photon index $\Gamma_{ph}\approx 3.8$.    
The accumulated by the present time {\it INTEGRAL} data allowed us to make, for the first time, 
the \textit{orbital phase-resolved X-ray spectroscopy}. 
Five orbital phase intervals chosen for spectral analysis 
are shown by the vertical dashed lines in Fig. \ref{f:lcbinned}. The obtained X-ray spectra 
within each orbital phase intervals I-V are shown in Fig. \ref{f:spectra}. It is seen
that within errors they have an identical power-law shape with the same 
photon spectral index 
$\Gamma_{ph}\simeq 3.8$. Nevertheless, there is 
a tendency of the spectrum to get
harder at $\phi_{orb}\sim 0.25$ (phase interval V in Fig. \ref{f:lcbinned}) than 
at $\phi_{orb}\sim 0.75$ (phase interval I in Fig. \ref{f:lcbinned}). 
Note also that at the middle of the 
eclipse (phase interval III in Fig. \ref{f:lcbinned}) the spectrum gets softer. This is exactly what is expected in the jet 
nutation picture: during the disk-jet nutation the angle between the line of sight
and the jet axis changes by $\sim 6$ degrees, so the observer looks into 
the funnel less (at $\phi_{orb}=0.5$) or more  (at $\phi_{orb}=0.75$) deeper, thus 
observing cooler or hotter parts of the jet base, respectively (see Section \ref{s_nut} below
for more detail). 
\textbf{The X-ray spectral variations across the orbital eclipse can provide insight into 
the nature and structure of the hard X-ray emission region (see, for example, discussion in 
\cite{Fil06} and \cite{Krivosheev09}). Therefore, 
observations with higher sensitivity are required to confirm or dismiss the spectral variations.}

\section{Nutational variability}
\label{s_nut}


From theoretical point of view, short-term motions in precessing
accretion disks in close binary systems occur when the binary
orbital period is not negligible compared with the precession
period. This is the case of SS 433. Theoretical analysis predicts
a $6^{d}.3$ periodic component of the moving spectral lines of SS
433 at an amplitude of $\sim 10\%$ of the $164^{d}$ precessional
motion \citep{Katz_ea82,LevineJernigan82,collins1986}, this
prediction is in a very good agreement with observations. Studies
by \citet{Katz_ea82,LevineJernigan82,collins1986} show that the
nutational motion in SS433 can be understood in terms of the forced
(slaved) disc precession \citep{heuvel1980,whitmire1980}.

Nutational variability in SS433 (nodding motion of tilted accretion disk) 
has been observed in the optical \citep{Goransk98, Davydov08} and radio
\citep{Trushkin_ea01} bands. 
At the time of the maximum disk opening (precessional phase 
zero, moment T3), the node line of the disk is normal to the line of sight, and
the nutational variability must be observed as a wobbling of the disk inclination angle 
to the line of sight with the nutational period $P_{nut}\approx 6.29$~days \citep{Katz_ea82, LevineJernigan82}. The analysis of an almost 30-years long  
optical spectroscopy of SS433 carried out by \cite{Davydov08} gives the 
ephemeris for the nutational motion of jets (the time when the $H_\alpha$
emission lines from jets are most widely separated)
$$
t_{max} = \hbox{JD} 2443009.720271 + (6^{\rm d}.287599  \pm 0^{\rm d}.00035  )\times E.
$$

A  significant flux excess at the orbital phase $\phi_{orb} \sim 0.25$ 
in the composite 18-60 kev {\it INTEGRAL} light curve (Fig. \ref{f:lc} and \ref{f:lcbinned}) 
is observed after the primary eclipse relative to the phase 0.75 before the eclipse. 
On the 40-60 keV light curve (Fig. \ref{f:hrorb}) 
two maxima with similar amplitude 
are clearly seen at both orbital phases 0.25 and 0.75. 
These $\sim 10\%$ 
sine-like variability occurring at about twice the orbital period, 
superimposed on the 
orbital light curve of SS433, is most likely due to the jet nutation.
The effect 
must be maximal at precessional phase zero when the 
binary system is observed in quadratures (i.e. at the binary phases 0.25 and 0.75). 
Due to nutation, the jet changes its inclination angle 
to the line of sight by $\delta\epsilon \approx 6^\circ$. 
Thus, the composite {\it INTEGRAL} 
light curve allowed us to see for the first time the jet nutation in the 40-60 keV band
\citep{Cher13}.

\begin{figure*}
\includegraphics[width=0.49\textwidth]{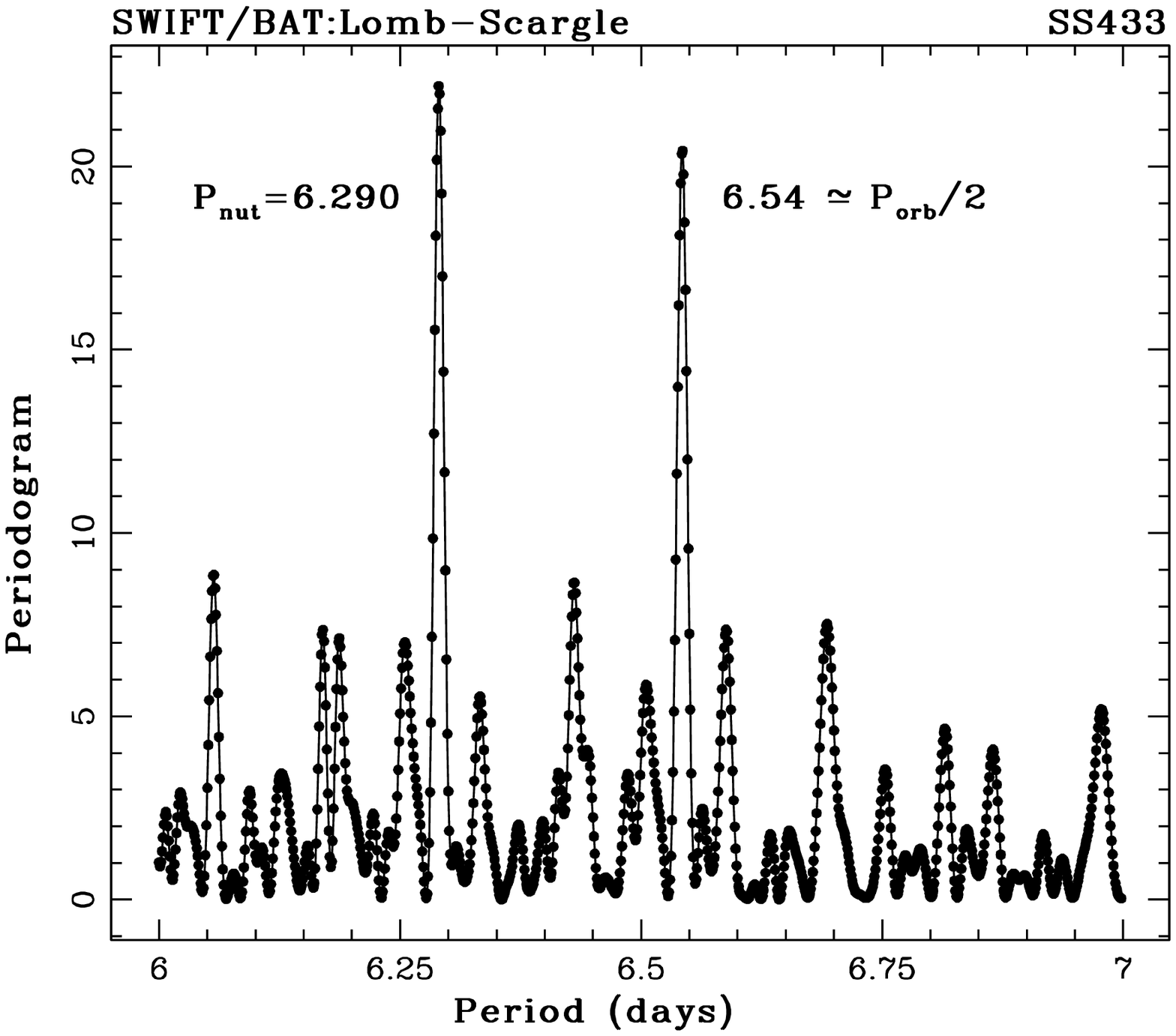}
\includegraphics[width=0.49\textwidth]{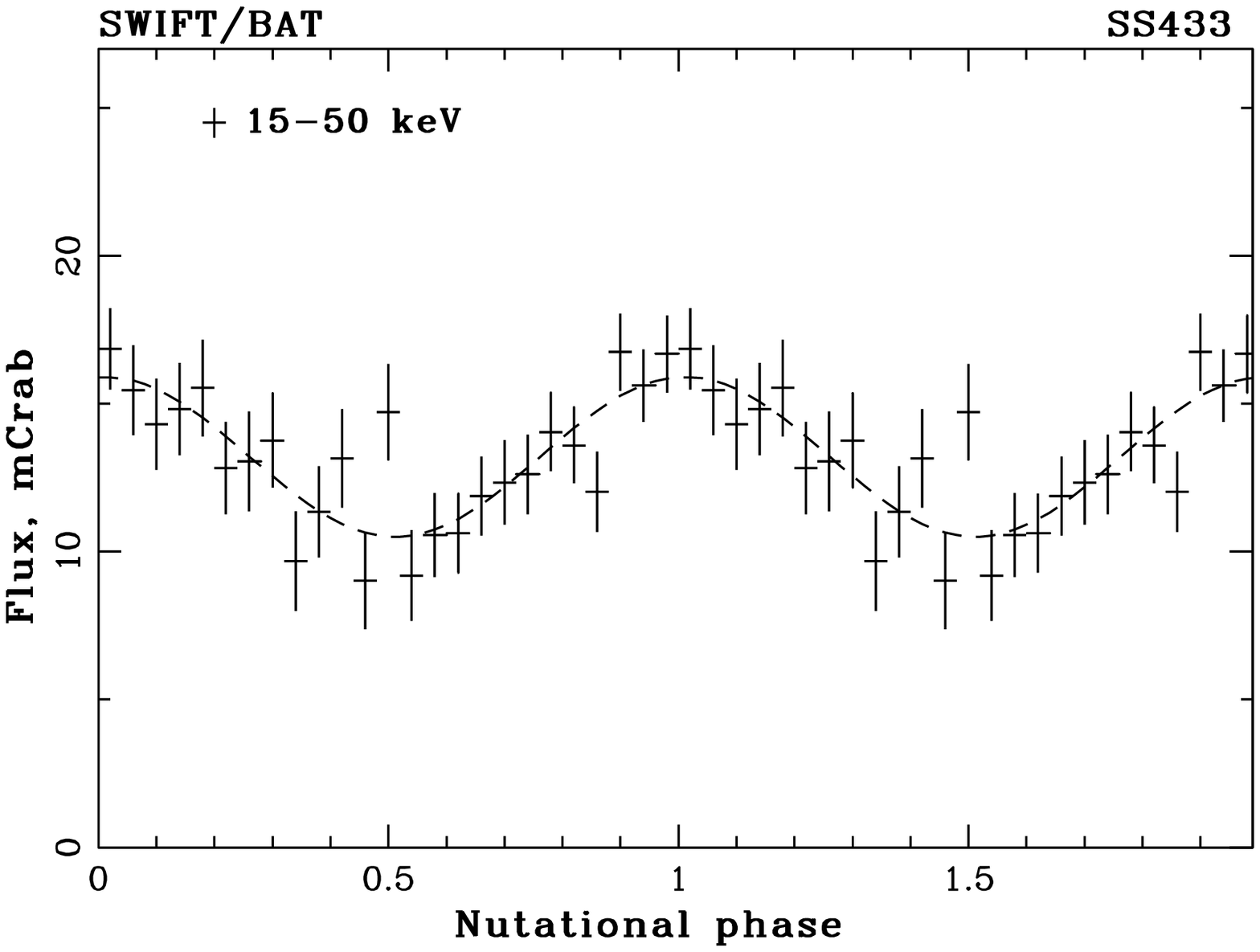}
\parbox[t]{0.49\textwidth}{\caption{The Lomb-Scargle periodogram constructed for \textit{Swift/BAT}
15-50 keV observations of SS433 at precessional phases $0.8<\psi_{pr}<1.2$. 
The peaks at about half orbital period $6^{\rm d}.54$ and nutation period
$6^{\rm d}.290$ are indicated.}\label{f:lsperiodogram}}
\hfill
\parbox[t]{0.49\textwidth}{\caption{Off-eclipse \textit{Swift/BAT} 15-50 keV nutation light curve of SS433. 
The dashed curve shows a sine-line best-fit to the data.}\label{f:swiftnut}}
\end{figure*}

\subsection{Nutational variability of SS433 from \textit{Swift/BAT} monitoring}

To substantiate this result, we first have performed the frequency analysis of 
the public archival data obtained with
the Burst Alert Telescope (BAT, 15-150 keV, Barthelmy et al.
2005) onboard the Swift observatory (Gehrels et al. 2004). 
For this purpose, we have used the 15-50 keV data seen only at precessional phases 
$0.8<\psi_{pr}<1.2$ (i.e. around moment T3). The 
Lomb-Scargle periodogram 
\citep{Lomb76, Scargle82} 
is shown in Fig. \ref{f:lsperiodogram}. The prominent
peaks at about half orbital period $6^{\rm d}.54$ and the nutation period
$6^{\rm d}.290$ are clearly seen. \textbf{Their amplitudes
imply a chance probability of less than $10^{-4}$ \citep{Baluev08}}. 
Note that the value of the nutation period
from the \textit{Swift/BAT} data agrees, within the errors, with that derived from 
the optical data \citep{Davydov08}. The \textit{Swift/BAT} nutational light curve
(only for observations near moments T3 $0.8<\psi_{pr}<1.2$ and 
off-eclipse with $0.8<\phi_{orb}<1.2$),
convolved with the $P_{nut}=6^{\rm d}.290$ period, is presented in Fig. \ref{f:swiftnut}.
\textbf{The folded \textit{Swift/BAT} lightcurve is not consistent with a
constant: the fit by a constant yields $\chi^2\approx 61$
for 25 d.o.f. Adding a sine component with period $6^{\rm d}.290$~d
strongly improves the fit, giving $\chi^2\approx 20$ for 24 d.o.f. 
We conclude that the sine line with amplitude
$\Delta F_{nut}=2.7\pm 0.5$~mCrab plus constant component 
$F_0=13.2\pm 0.4$~mCrab (the dashed line
in Fig. \ref{f:swiftnut}) adequately describes the data.}
The relative \textit{Swift/BAT} nutational variability 
is $\Delta F_{nut}/F_0\approx 0.205$.

\subsection{Fitting nutational variability in the \textit{INTEGRAL} orbital light curve}

Following
\citet{Katz_ea82, LevineJernigan82, collins1986} we parametrize
the change in the observed flux $\Delta F$ due to the 
nutational motion of the disk and jets 
as follows:
\begin{equation}
\Delta F_{nut}=A\sin(\omega_{nut}t+c_0)=a \sin
(2\pi(\varphi_{orb}+\varphi_0)P_{orb}/P_{nut}), \label{nutvar}
\end{equation}
where $A$ is the amplitude,
$\omega_{nut}=2\pi/P_{nut}$ is the nutational angular velocity,
$t$ is the time, $c_0$ is a constant,
$\varphi_{orb}=t/P_{orb}$ is the orbital phase, $\varphi_0$ is the
phase angle between the orbital and the nutational minima,
$P_{orb}$ is the orbital period of the system, $P_{nut}$ is the
period of nutation. The ratio $P_{orb}/P_{nut}=2.08$ is fixed,
while $A$ and $\varphi_0$ are free parameters (see e.g. \citet{Katz_ea82}, equation [20]). 
The third mode of nutational variability is the dominant $6^{d}.3$
mode, its angular velocity is $\omega_{nut}$=$2\dot\eta-\Omega_s$,
where $\eta$ is the orbital frequency of the companion star,
$\Omega_s$ is the mean precessional rate of the accretion ring. We
neglect the term $0.5\dot\Omega t^2$ in those equation, because
all our data are averaged over one orbital period. 

\begin{figure*}
\includegraphics[width=0.47\textwidth]{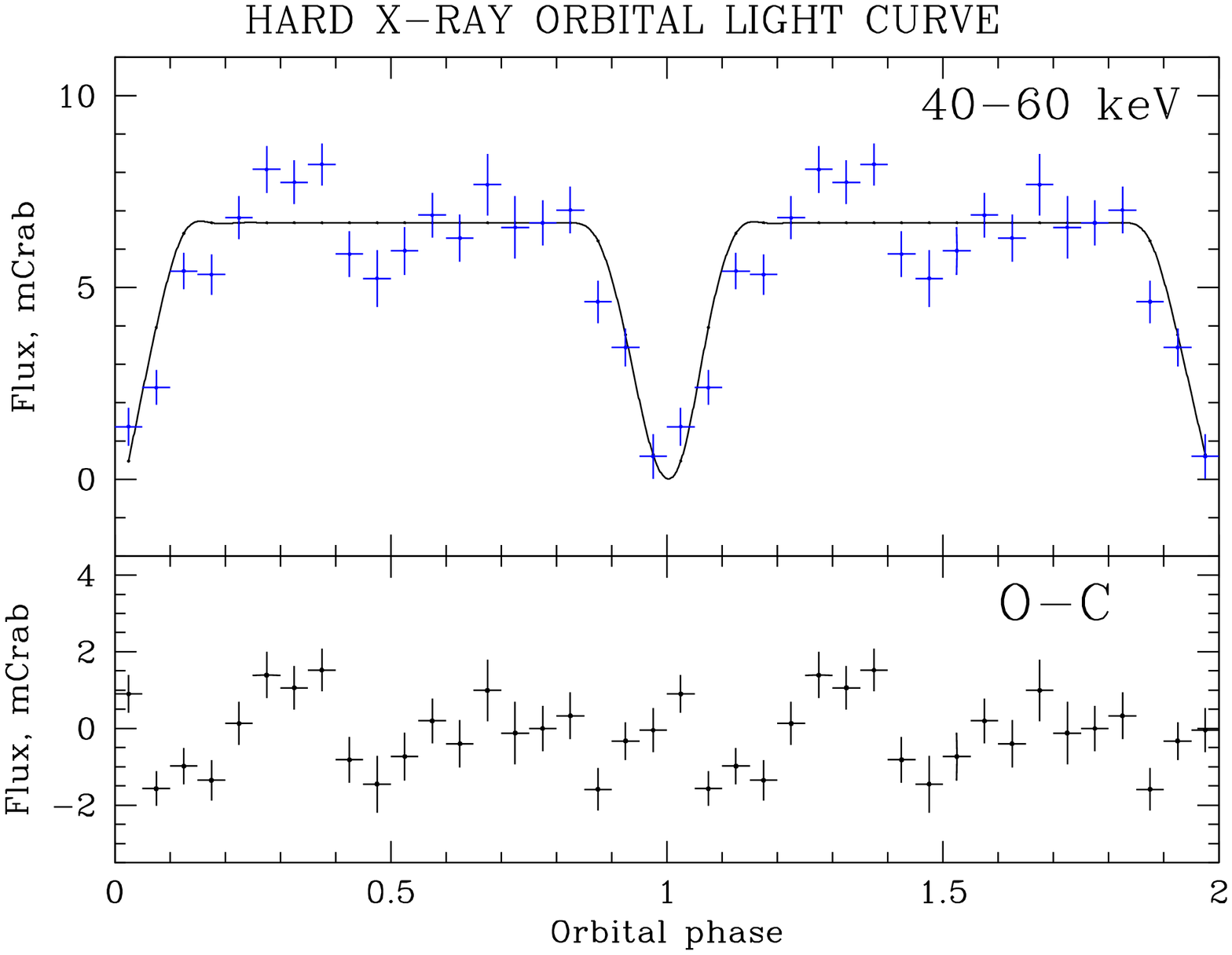}
\includegraphics[width=0.47\textwidth]{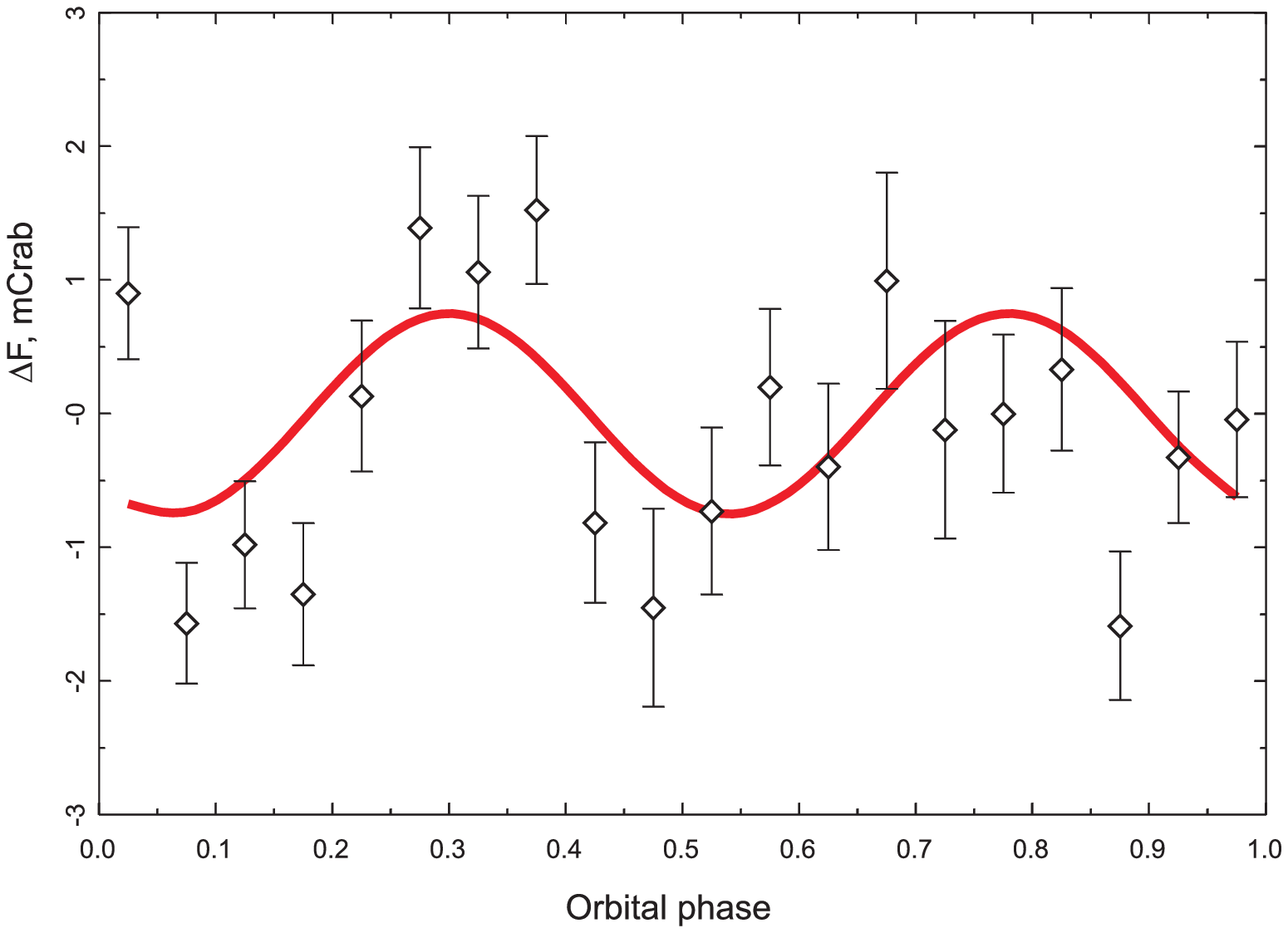}
\parbox[t]{0.47\textwidth}{\caption{40-60 keV light curve fitted 
with the geometrical model of the primary eclipse for $q=0.3$ (upper panel)
and the residuals (bottom panel).}\label{f:nutres}}
\hfill
\parbox[t]{0.47\textwidth}{\caption{Residuals of the 40-60 keV orbital light curve fitted 
with nutational variability of SS433 (equation (1)).}\label{f:nutresfit}}
\end{figure*}

We have performed an independent analysis of the nutational variability 
in the {\it INTEGRAL} 40-60 keV data. To this aim, we have subtracted
the best-fit geometrical primary eclipse light curve calculated for
the binary mass ratio $q=0.3$ (see Section \ref{s:param} for more detail)
(Fig. \ref{f:nutres}, the solid curve on the upper panel). 
The residual light curve (points with error bars in Fig. \ref{f:nutresfit})  has been fitted
with a sine curve (\ref{nutvar}) (the solid line in Fig. \ref{f:nutresfit}) 
with parameters $A=0.75$~mCrab, $\varphi_0=0.30$. The
value of $\chi^2$-test for this fit is $\chi^2=47.63$ for 18 d.o.f., giving 
the reduced $\chi^2_{dof}\approx 2.65$.
With such an amplitude and
$F_0\approx 7$~mCrab (see Fig. \ref{f:nutres}),
the fractional 40-60 keV flux change due to the nutational motion is 
$\Delta F_{nut}/F_0\approx 0.11$, which is smaller but consistent, within errors, with the 
\textit{Swift/BAT} amplitude $\approx 0.205$. 
The larger value of $\chi^2_{dof}\approx 2.65$ for \textit{INTEGRAL}
data in comparison with \textit{Swift/BAT} nutational light curve fitting ($\chi^2_{dof}\approx 1$)
is due to our using, in this Section, data folded with the orbital period of SS433; as $P_{orb}/2$ and $P_{nut}$ are slightly different, the dispersion of points around the mean nutational
light curve in the \textit{INTEGRAL} data are larger than for the \textit{Swift/BAT} data.

\subsection{Geometrical model of nutational variability in SS433}

\begin{figure*}
\centering{\includegraphics[width=0.4\textwidth]{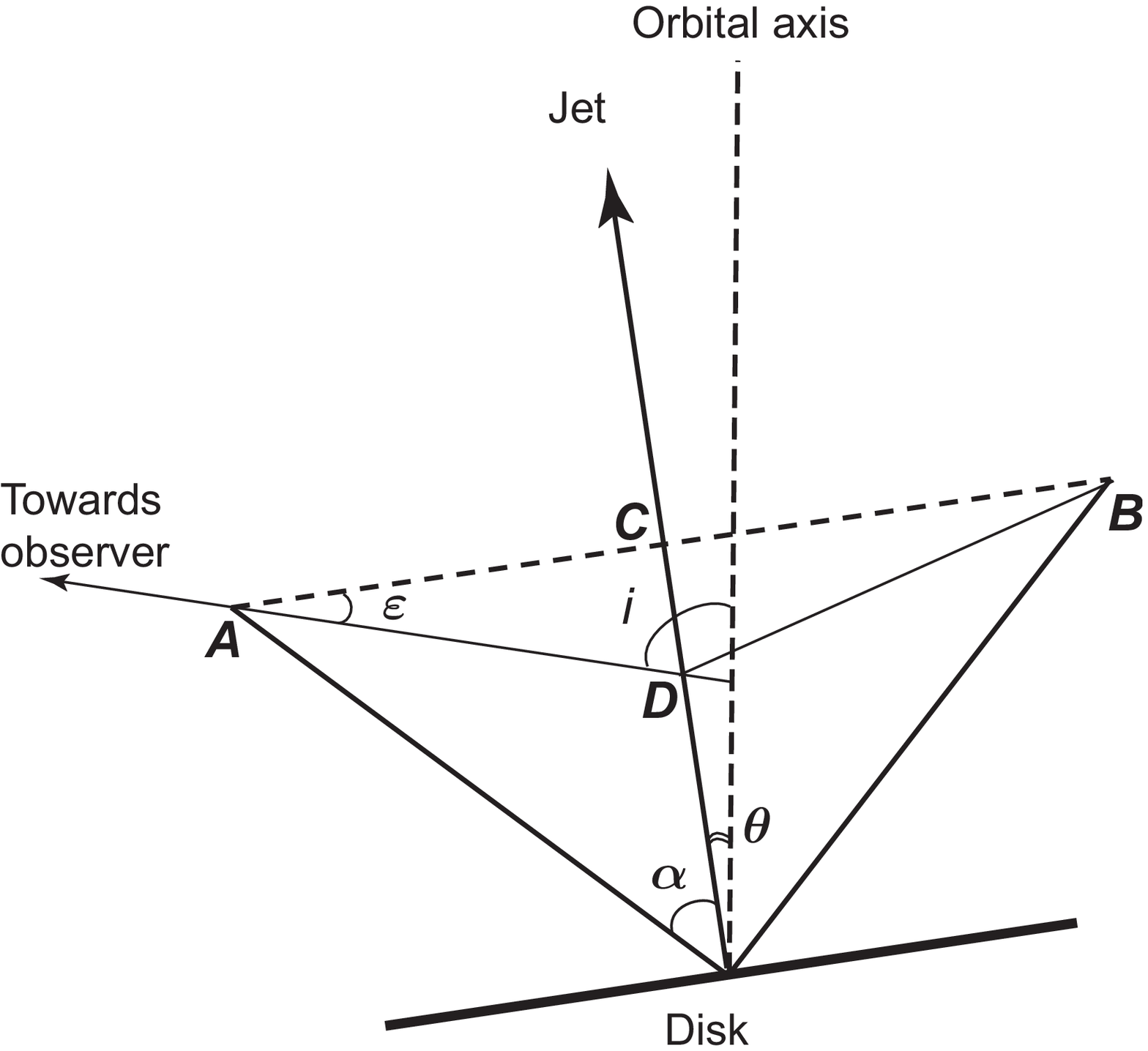}}
\centering{\includegraphics[width=0.4\textwidth]{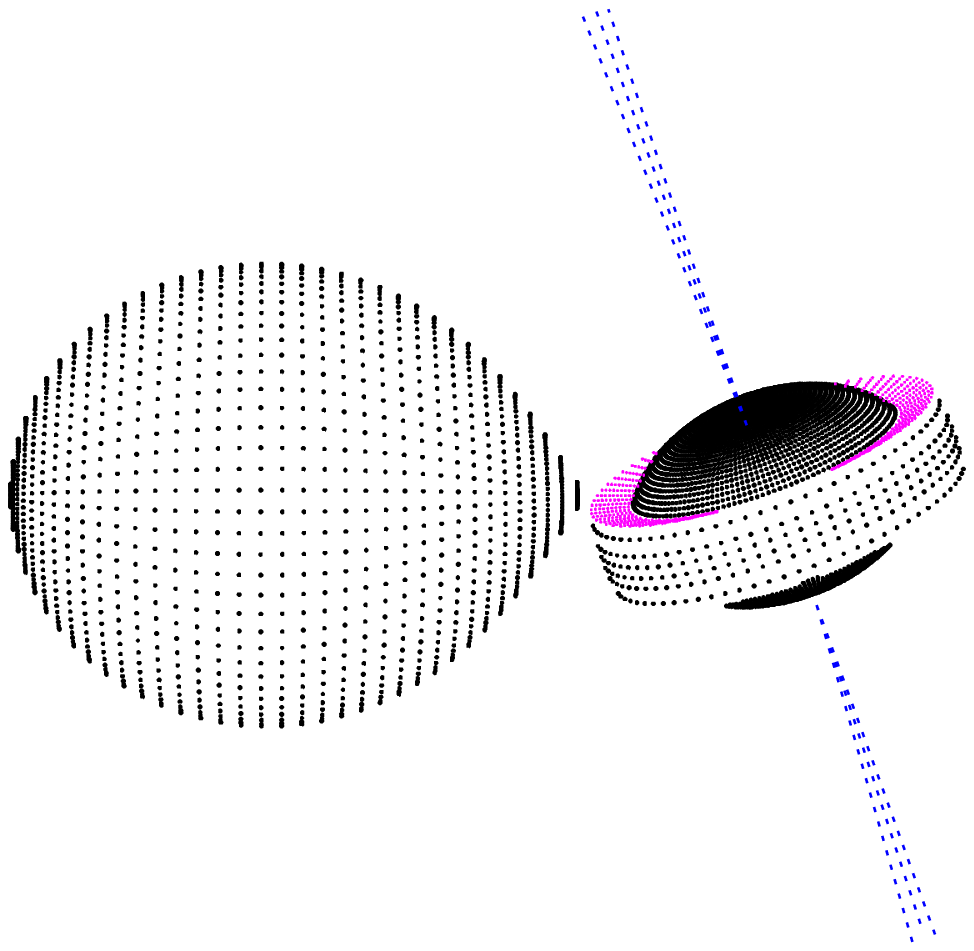}}
\parbox[t]{0.47\textwidth}{\caption{Scheme of nutation angles in SS433. Binary inclination angle is $i=79^\circ$, 
jet inclination angle to the orbital angular momentum is $\theta=20^\circ$, 
jet funnel opening half-angle is $\alpha<(i-\theta)$.}\label{f:nutangles}}
\hfill
\parbox[t]{0.47\textwidth}{\caption{
Model view of SS433 for binary mass ratio $q=0.3$ is shown at the 
disk precession angle 80 degrees. Thin jets normal to the disk plane 
are shown, but their contribution to hard X-ray emission is ignored. 
}\label{f:model}}
\end{figure*}

The relative nutational flux variability $10-15\%$ can be explained in 
the frame of the simplest geometrical model of nodding funnel
around jets filled with hot scattering plasma. Let the line
of sight make an 
angle $\epsilon$ with the jet normal, as sketched in Fig. \ref{f:nutangles}. 
From the kinematical model of SS433 \citep{Margon84} we adopt  
the binary inclination angle to be $i\approx79^\circ$, the jet precession angle
to be $\theta\approx 20^\circ$, and the jet nutation angle to be $\delta \theta\approx 6^\circ$. 
We also assume the opening angle of the jet funnel $\alpha$. 
From the geometry (Fig. \ref{f:nutangles}) we find $\epsilon=\pi/2-i+\theta$.  
If $\alpha<i-\theta=59^\circ$ (which is supported by 
numerical simulations of wind from supercritical accretion disk in SS433,
see e.g. \cite{Okuda_ea09}), then the relative change in the volume of the
jet funnel filled with hot scattering plasma  
seen by the observer due to jet nutation is
$$
\frac{\delta V}{V}=\frac{\delta CD}{CD}=\frac{\delta \tan \epsilon}{\tan \epsilon}=
\frac{\delta \theta}{\cos{(i-\theta)}\sin{(i-\theta)}}\approx 0.23
$$
(here the observed volume is approximated by the cone ABD in Fig. \ref{f:nutangles};
the closer the jet funnel angle $\alpha$ to $(i-\theta)$, the better this approximation). 
Therefore, in the first approximation, the relative 
change in the hard X-ray flux from this volume in one nutational period is 
$\delta F/F=\delta V/V\approx 0.23$, which is somewhat larger than the observed 
value $0.11\div 0.20$ obtained in our analysis of the \textit{INTEGRAL} and \textit{Swift/BAT} data. 
However, in view of roughness of the
simple geometrical model, we can conclude that the observed relative amplitude of
the nutational variability is in agreement 
with the geometry of jets and accretion funnel in SS433.

\section{Parameters of SS433 from analysis of hard X-ray light curves}
\label{s:param}

\subsection{Broadband 18-60 keV orbital light curve}

\begin{figure*}
\includegraphics[width=0.3\textwidth]{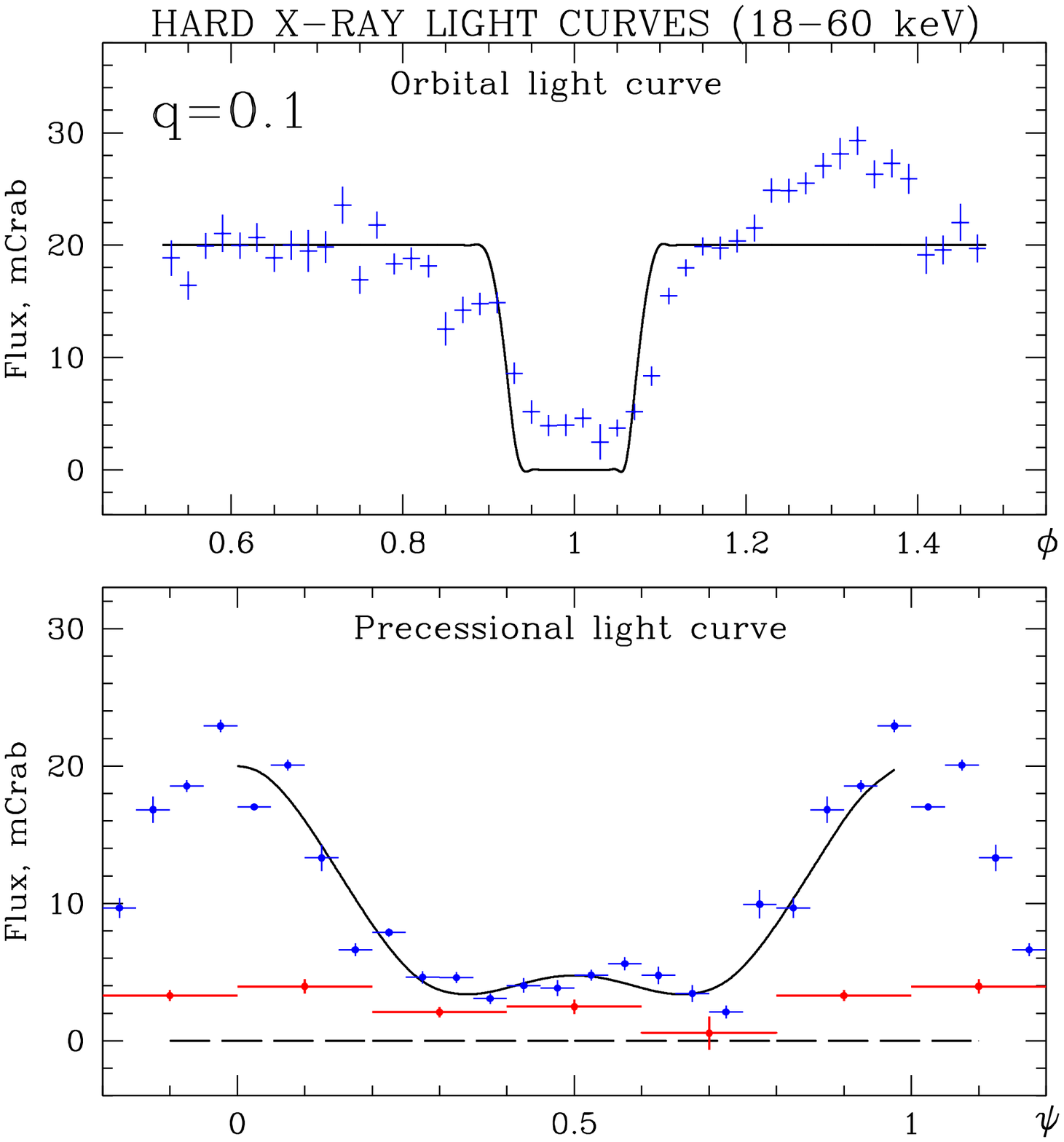} 
\hfill
\includegraphics[width=0.3\textwidth]{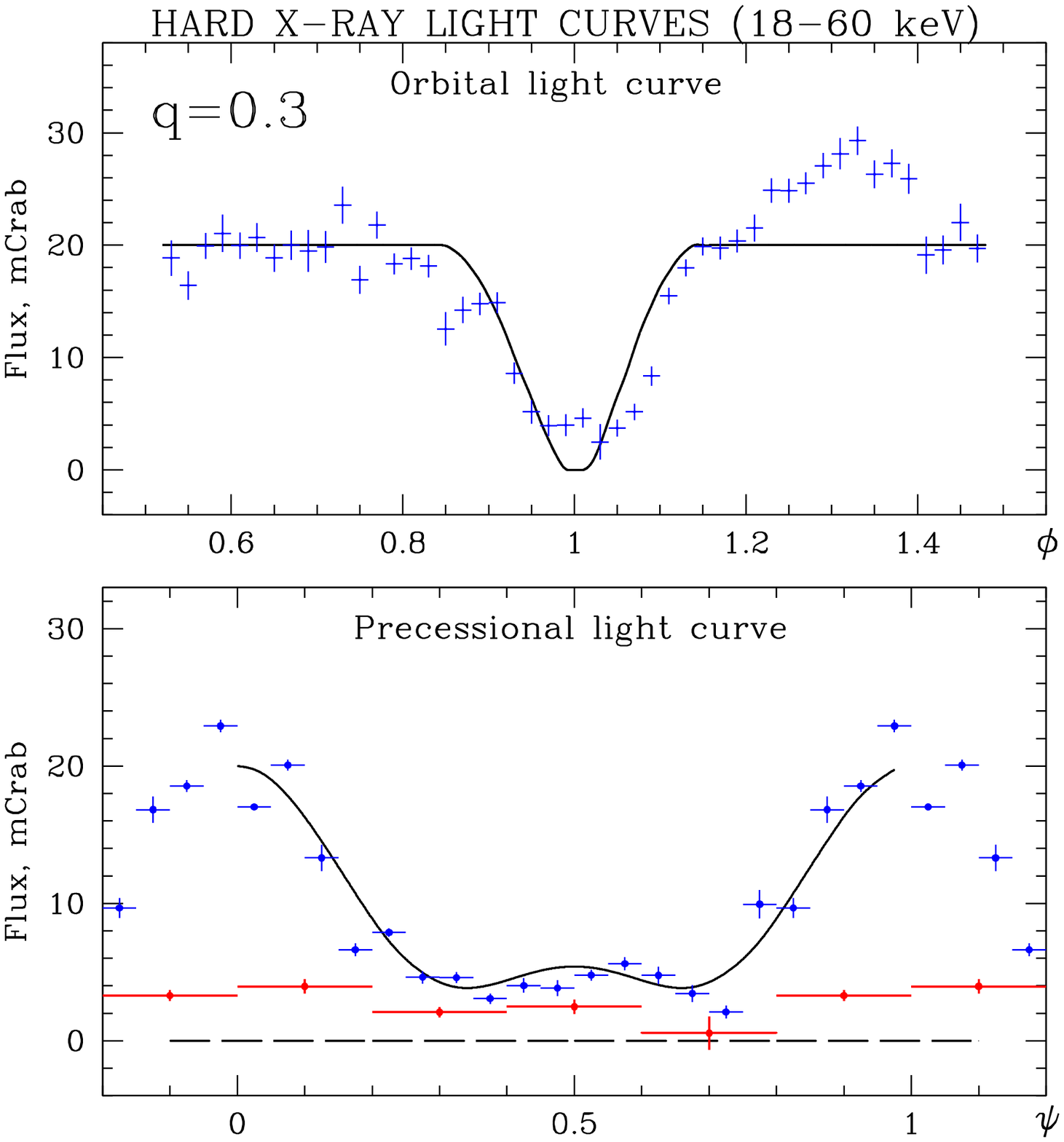} 
\hfill
\includegraphics[width=0.3\textwidth]{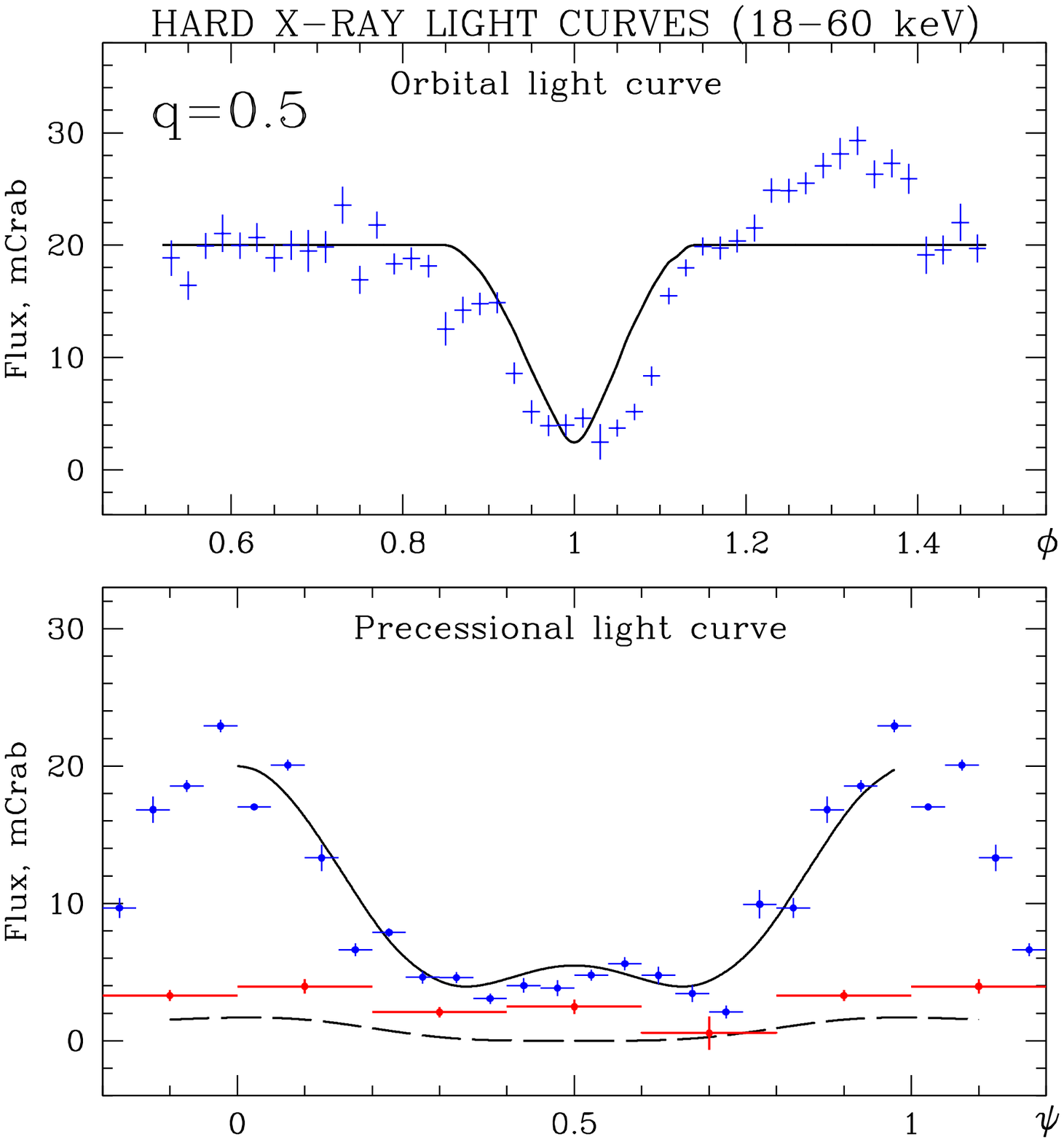}
\caption{Joint analysis of orbital (upper plots) and precessional (lower plots) 
18-60 keV light curves of SS433. 
On the upper plots, the model orbital light curve is shown by the solid line. 
On the bottom panels, the model precessional 
light curves off eclipses ($0.2<\phi_{orb}<0.8$) and in eclipses ($0.95<\phi_{orb}<1.05$)  are shown by the solid and dashed lines, respectively. 
The observed 
precessional variability in the middle of eclipse are shown by the gray (red in color on-line version) crosses. 
Left panel: $q=0.1$, 'short jet' corona
($a_j=0.25$, $b_j=0.1$, 
$\alpha=40^o$). 
Middle panel$q=0.3$, 'short jet' corona
($a_j=0.35$, $b_j=0.13$, 
$\alpha=80^o$). 
Right panel: $q=0.5$, 'short jet' corona
($a_j=0.35$, $b_j=0.13$, 
$\alpha=80^o$). }
\label{f:lcanalysis1860}
\end{figure*}

The {\it INTEGRAL} observations of SS433 provide three different light curves, which can be used to constrain the parameters of the system: 1) the orbital light curve, 2) the precessional light
curve out of eclipses, and 3) precessional light curve in the middle of the eclipses (see 
Fig. \ref{f:lcanalysis1860}, upper panels, and lower panels). 
We use a geometrical model described in detail in \citet{Ant92, Cher09}. Briefly, 
the model includes an optical star with mass $M_v$ filling its Roche lobe, 
a compact star with mass $M_x$ surrounded by an optically thick 
accretion disk with radius $a_d$, 
and a hot corona
which is modeled as a broad 'jet' parametrized by the part of ellipse 
with semi-axes $a_j$ and $b_j$ normalized to the binary orbital separation. The elliptical corona 
is restricted by the cone with half-angle $\alpha$ (see Fig. \ref{f:model}).    
The shape and amplitude
of the precessional light curve constrain the height of the hot X-ray corona, while  
the orbital light curve restricts the accretion disk radius. 
At a given binary mass ratio, 
after finding the best parameters for the precessional variability, we 
can calculate the deviations of the model orbital light curve from the observed one.  
The results of the joint analysis of the orbital and precessional 18-60 keV 
light curves of SS433 are shown in Fig. \ref{f:lcanalysis1860}. In the present analysis,
we neglected the contribution of thermal X-ray emission from jets. 

\textbf{The
new \textit{INTEGRAL} data added after 2008 fully confirm the results of our analysis 
published in \cite{Cher09}. While the eclisping light curve alone
can be well reproduced by a small mass ratio $q=m_x/m_v\sim 0.1$ in the model
with 'long' X-ray emitting jet ($b_j>0.5$), the observed precessional 
light curve cannot be reproduced by models with 
small binary mass ratios $q\le 0.25$ (see Fig. \ref{f:lcanalysis}a below). Models with 'short'
X-ray emitting jet ($b_j\sim 0.1\div 0.2$) and small mass ratio $q=0.1$ can 
fit the observed precessional variability, but fail in describing the X-ray primary eclipse form (see Fig. \ref{f:lcanalysis1860}a).}

\begin{figure*}
\includegraphics[width=0.5\textwidth]{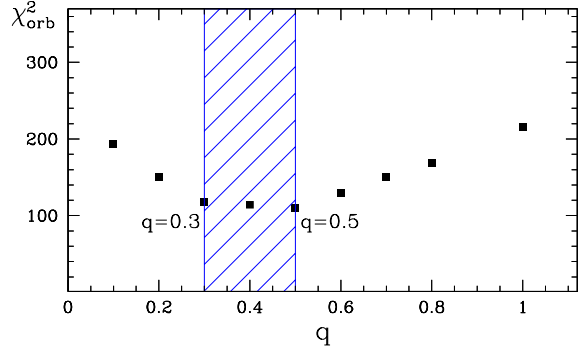}
\caption{ $\chi^2$-value for the orbital light curve (18-60 keV) for different binary mass ratios $q$.}
\label{f:lcchi2}
\end{figure*}

\begin{figure*}
\includegraphics[width=0.3\textwidth]{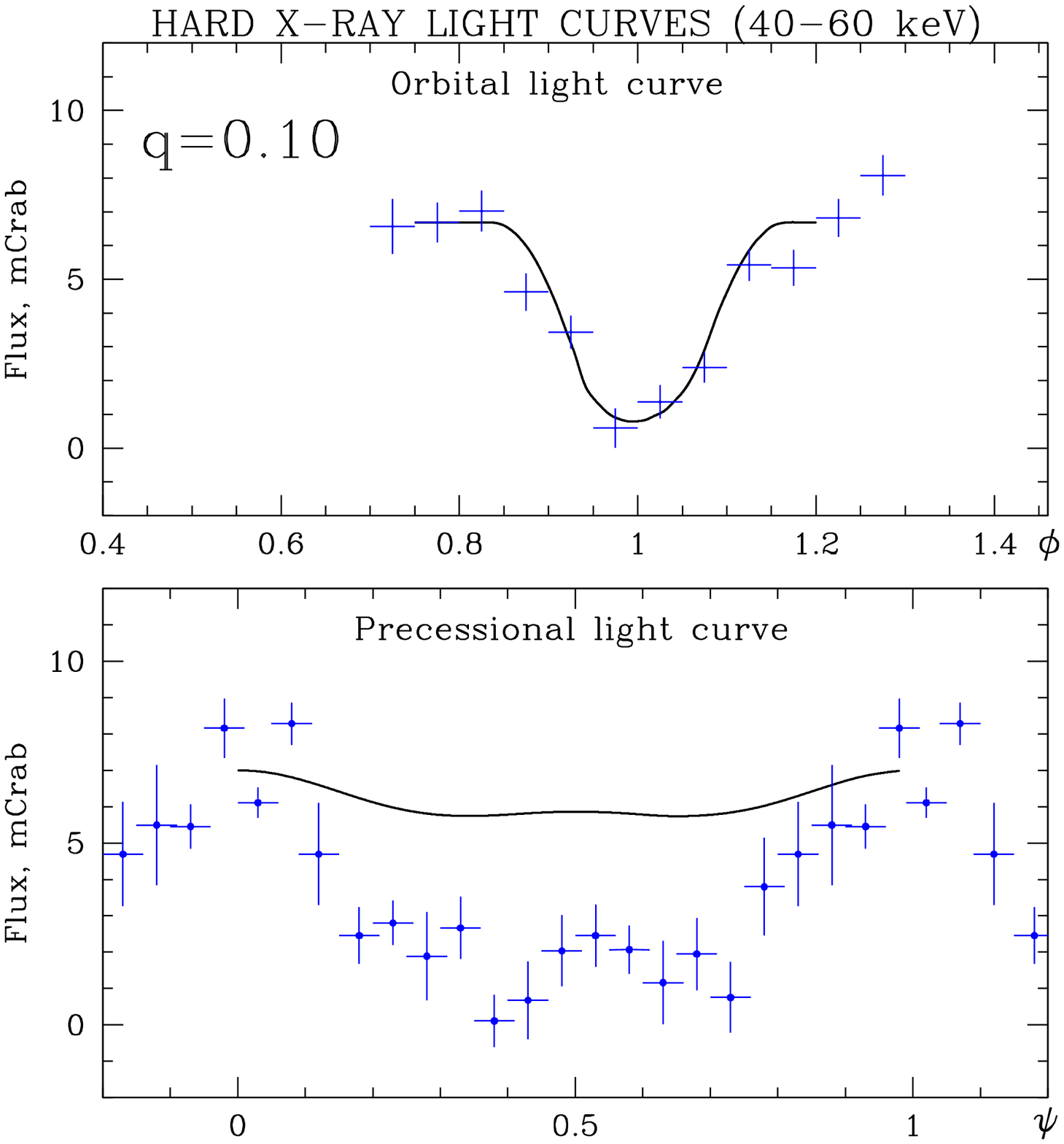}
\hfill
\includegraphics[width=0.3\textwidth]{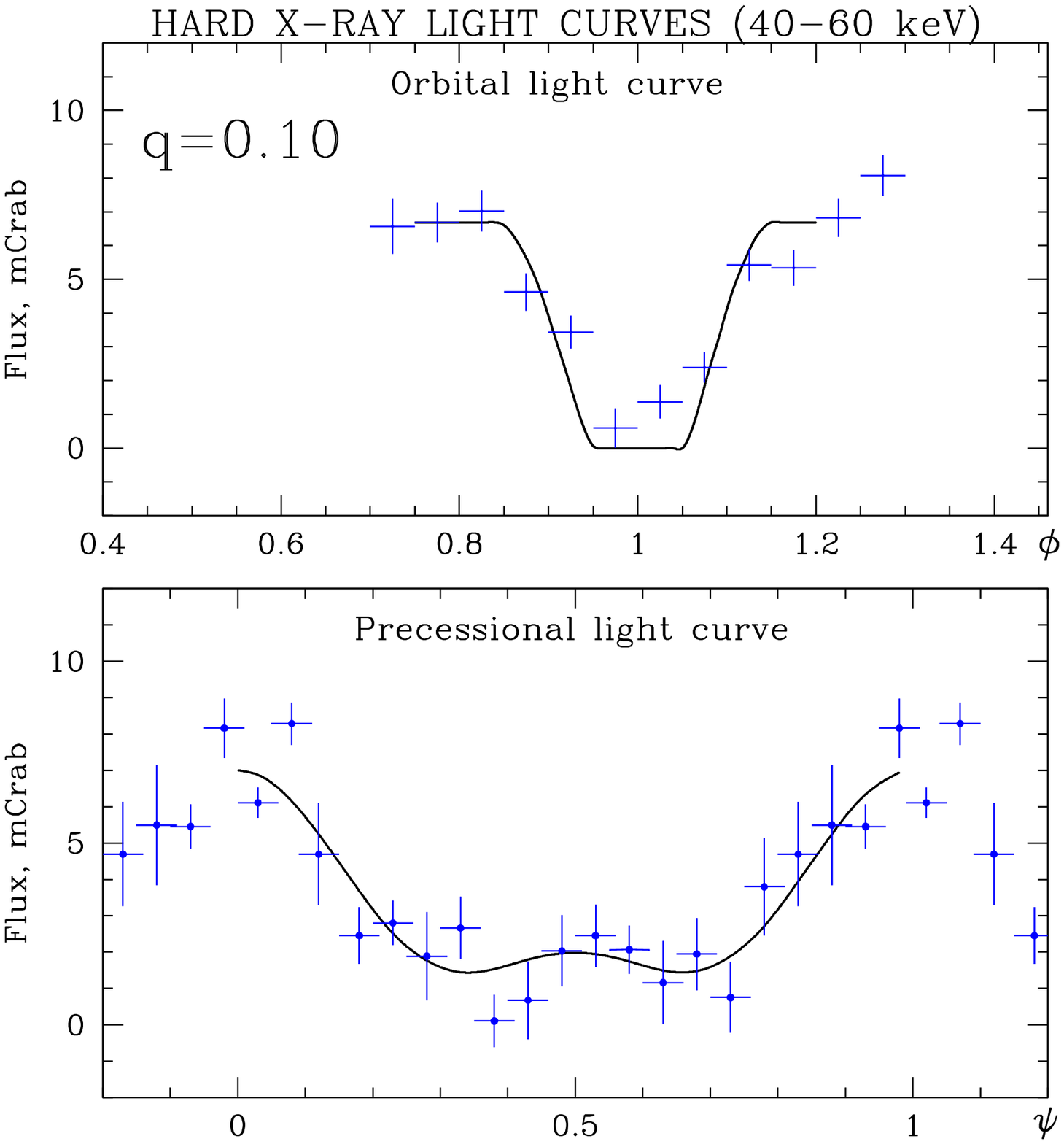} 
\hfill
\includegraphics[width=0.3\textwidth]{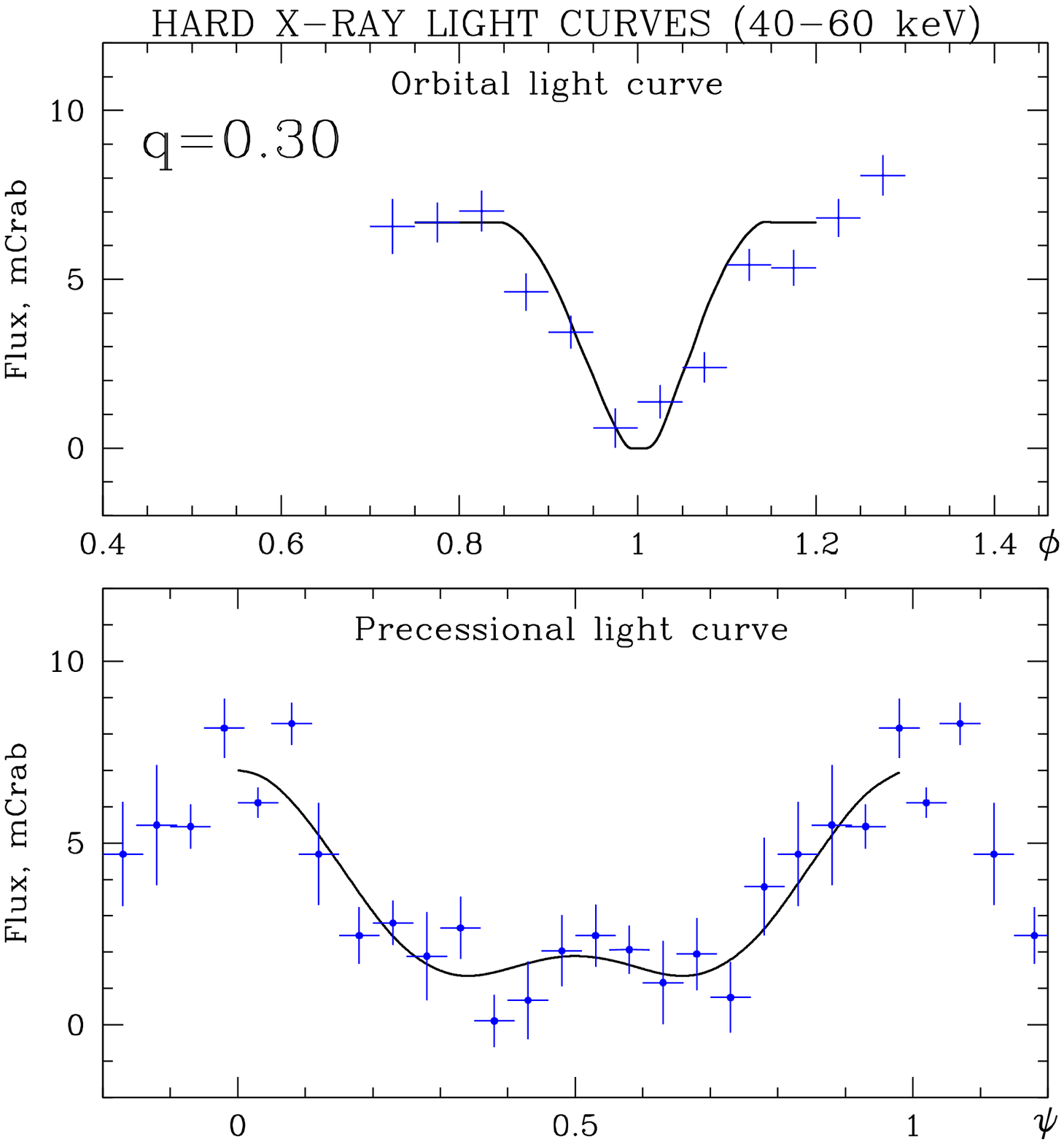} 
\caption{Joint analysis of orbital (upper plots) and precessional (lower plots) 
40-60 keV IBIS/ISGRI light curves of SS433. Left panel: $q=0.1$, 'long jet' corona ($a_j=0.25$, $b_j=0.55$, 
$\alpha=80^o$); only orbital 
light curve can be reproduced. Middle panel: $q=0.1$, 'short jet' corona
($a_j=0.25$, $b_j=0.1$, 
$\alpha=80^o$). Both 
orbital and precessional light curvs can be fitted, but the total eclipse (plateau at zero flux) appears (unobserved). Right panel: $q=0.3$, 'short jet' corona
($a_j=0.35$, $b_j=0.13$, 
$\alpha=80^o$); both orbital and precessional light curves are well reproduced.}
\label{f:lcanalysis}
\end{figure*}

In Fig. \ref{f:lcchi2} we plot the values of $\chi^2$ for 
orbital 18-60 keV light curve of SS433 \textbf{shown in upper panels of Fig. \ref{f:lcanalysis1860}} as a function of 
the binary mass ratio $q$ (\textbf{i.e. with the disk and coronal 'jet' parameters which 
fit the precessional ligth curve as shown in lower pannels of Fig. \ref{f:lcanalysis1860}}). It is seen that within the
interval $0.3<q<0.5$ the model fits yield similar $\chi^2$ values. 
This mass ratio range is
consistent with results of an independent 
analysis of the orbital and precessional variability of SS433 in the optical \citep{AntCher87}.


\subsection{Hard X-ray (40-60 keV) orbital light curve}

To avoid contamination from thermal X-ray emission from relativistic jets, 
we repeated the analysis using only hard X-ray 40-60 keV light curves 
shown in Fig. \ref{f:hprec} and Fig. \ref{f:hrorb}. 
The result for different mass ratios $q=M_x/M_v$ is presented in  
Fig. \ref{f:lcanalysis}. It is seen that the hard
X-ray orbital and precessional light curves can be simultaneously reproduced 
by the geometrical model for the binary mass ratio $q\gtrsim 0.3$ (at smaller mass ratios a
plateau corresponding to the total eclipse of the hot corona by the optical star appears 
in the orbital light curve, which is not observed), in agreement with the analysis of
the broadband 18-60 keV orbital and precessional light curves shown in Fig. \ref{f:lcanalysis1860}.

\section{Discussion}
\label{s_disc}

The analysis of all {\it INTEGRAL} observations of SS433  in
2003-2011 confirms our earlier conclusions about binary system
parameters \citep{Cher09}. With the obtained range of the binary
mass ratio $q=0.3\div 0.5$, there are three possible ways to estimate the
mass of the compact star $m_x$ in SS433.

\begin{enumerate}

\item Using optical spectroscopy of the optical component A7I in
the range $ 4400 \div 5000$~A,  \cite{Hillwig_ea04} and \cite{hillwig08}
measured the semiamplitude of the radial velocity curve of the mass donor star of
$58.2\pm 3.1$ km s$^{-1}$, yielding the optical star mass function
\begin{equation}
f_v(m)=\frac{m_x\sin^3i}{(1+1/q)^2}\simeq 0.268 M_\odot ,
\end{equation}
For $q = 0.3\div0.5$, from this mass function we obtain masses of the 
compact object and the optical star 
$m_x\simeq 5.3 M_\odot \div 2.6 M_\odot$ and $m_v\simeq 17.7 M_\odot \div 5.1 M_\odot$, respectively.  

\item From the optical spectroscopy of the accretion disk
emission lines \citep{crampton1981,fabrika1990} (using He II
4686 A 
emission line) and \citet{gies2002} (using CII
7231, 7236 A emission lines) derived the semiamplitude for
the compact star radial velocity $195$ km s$^{-1}$ \citep{crampton1981}, $175$
km s$^{-1}$ \citep{fabrika1990}, $162$ km s$^{-1}$
\citep{gies2002}. The average value of the compact star mass
function for the last two compact star radial velocity estimates is
\begin{equation}
f_x(m)=\frac{m_v\sin^3i}{(1+q)^2}\simeq 6.4 M_\odot ,
\end{equation}
\noindent For $q = 0.3\div 0.5$ and the orbital inclination  angle
$i\simeq 78.8^{\circ}$ \citep{MargonAnderson89}, this yields
masses of the compact object and the optical star in SS433 $
m_x\simeq 3.44 M_\odot \div 7.63 M_\odot$, $m_v\simeq 11.46 M_\odot \div 15.26 M_\odot$, respectively.

\item From the high-resolution optical spectroscopy of circumbinary
shell of SS433 \cite{blundell2008} and \cite{Bowler13} estimated the total mass of
SS433 binary system to be $m_x+m_v\gtrsim 40 M_\odot$. For $q=0.3\div 0.5$
this gives the lower limits of the masses of the components in SS433: 
$ m_x\gtrsim 9.23 M_\odot \div 13.33 M_\odot $, $m_v\gtrsim 30.77 M_\odot \div 26.67 M_\odot$.
\textbf{The compact object mass derived from these observations is about twice as high
as from radial velocity measurements.}

\end{enumerate}

\textbf{Therefore, three independent estimates of the compact
star mass in SS433 (higher than $\approx 2.6\div 3 M_\odot$) suggest it to be a black hole.
However, for accurate determination of the mass of the compact star, 
further high-resolution spectroscopic observations of SS433 
are highly desirable to improve estimates of the spectral class and radial velocity curve of the optical star. It is also important to improve the radial velocity curve from
observations of emission lines (He II 4686, CII 7231,7236, etc.) in order to reliably 
determine the mass function of the compact star.}

\section{Conclusions}
\label{s_concl}

1) {\it INTEGRAL} observations of SS433 in hard X-rays 18-60 keV allowed us for the first time 
to make orbital-resolved spectroscopy of the X-ray eclipse in the precessional phase
corresponding to the maximum opening angle of the disk. The hard X-ray continuum is fitted with a
power-law with photon index $\Gamma\approx 3.8$ which does not 
significantly change across the eclipse, 
suggesting the origin of this emission as being due to scattering in hot corona 
surrounding the funnel around the jets in a supercritical accretion disk in SS433. 

2) For the first time, {\it INTEGRAL} observations of SS433 revealed the presense of the secondary 
maximum in the hard X-ray precession light curve of the source at the precessional
phase $\psi_{pr}\approx 0.6$ of the relative amplitude $\sim 2$ (with a total 
relative amplitude of the precessional variability of 5-7). The secondary maximum is
seen in both broad-band 18-60 keV and hardest 40-60 keV bands.    

3) For the first time, the joint analysis of hard X-ray (40-60 keV) orbital and precessional light curves has been performed. This analysis independently 
confirms our previous result \citep{Cher09} that 
the low value of the mass ratio $q=M_x/M_v$ in SS433 cannot reproduce the observed orbital
and precessional light curves. With the existing estimates of the 
mass function of the optical and compact star, the \textbf{obtained binary mass ratio range 
$q\sim 0.3\div 0.5$} 
points to the black hole nature of the compact star in SS433. The black-hole mass of
the compact star is found from three independent measurements. 

4) The shape of the hard X-ray orbital light curve 18-60 keV demonstrates two humps at
around orbital phases 0.25 and 0.75, most likely due to the nutation effects in 
SS433. These humps are mostly pronounced in the 40-60 keV orbital light curve. 
The nutation effect in SS433 with a period of $P_{nut}\simeq 6^d.290$ with similar relative amplitude 
is independently confirmed by the analysis of \textit{Swift/BAT} observations of SS433.

New {\it INTEGRAL} observations of SS433 at different precessional phases will
be used to further constrain physical parameters of this unique galactic microquasar.

\section*{Acknowledgments}
\textbf{We thank the anonymous referee for critical reading of the manuscript 
and useful notes which allowed us to improve the paper.
The results of this work are based
on observations with \textit{INTEGRAL}, an ESA project with instruments and science data
centre funded by ESA member states (especially the PI countries: Denmark, France, Germany,
Italy, Switzerland, Spain), and Poland, and with the participation of Russia and the USA.
The data were obtained from the European and Russian
\textit{INTEGRAL} Science Data Centers{\footnote{http://isdc.unige.ch
}$^,$\footnote{http://hea.iki.rssi.ru/rsdc
}}. The work of AMCh, EAA and AIB was 
partially supported by the RFBR grant 11-02-00258 and 
by the Program for State Support of Leading Scientific Schools of RF
2374.2012.2. SVM was partially supported by RFBR grant 11-02-01328 and 
the Ministry of Education and Science of RF through Contract N8701. KAP was 
partially supported by RFBR grant 12-02-00186.}


\begin{thebibliography}{99}

\bibitem[\protect\citeauthoryear{Antokhina \& Cherepashchuk}{1987}]{AntCher87}
Antokhina E.A., Cherepashchuk A.M., 1987, SvA, 31, 295

\bibitem[\protect\citeauthoryear{Antokhina et al.}{1992}]{Ant92}
Antokhina E.A., Seifina E.V., 
Cherepashchuk A.M., 1992, SvA, 36, 143

\bibitem[\protect\citeauthoryear{Baluev}{2008}]{Baluev08}
Baluev R.V., 2008, MNRAS, 385, 1279


\bibitem[\protect\citeauthoryear{Barthelmy et al.}{2005}]{BAT}
Barthelmy S.D. et al., 2005, Space Sci. Rev., 120, 143

\bibitem[\protect\citeauthoryear{Begelman et al.}{2006}]{Begelman06}
Begelman M.C., King A.R., Pringle J.E., 2006, MNRAS, 370, 399

\bibitem[\protect\citeauthoryear{Blundell et al.}{2008}]{blundell2008} Blundell K. M., Bowler M. G., Schmidtobreick L., 2008, ApJ, 678, L47

\bibitem[\protect\citeauthoryear{Bowler}{2013}]{Bowler13} 
Bowler M. G., 2013, A\&A, 556, A149 

\bibitem[\protect\citeauthoryear{Brinkman et al.}{1989}]{brinkman89}
Brinkman W., Kawai N., Matsuoka M., 1989, A\&A, 218, L13

\bibitem[\protect\citeauthoryear{Burenin et al.}{2010}]{Burenin2010}
Burenin R.A., Revnivtsev M.G., Khamitov I.M., 2011, Astron. Letters, 37, 100

\bibitem[\protect\citeauthoryear{Cherepashchuk}{1981}]{Cher81}
Cherepashchuk A.M., 1981, MNRAS, 194, 761

\bibitem[\protect\citeauthoryear{Cherepashchuk}{1989}]{Cher88}
Cherepashchuk A.M., 1989, Sov. Sci. Rev. Ap. Space Phys., Ed. by R.A.Sunyaev, v.7, p.185

\bibitem[\protect\citeauthoryear{Cherepashchuk et al.}{1995}]{Cher95}
Cherepashchuk A.M., Bychkov K.V., Seifina E.V., 
1995, ApSS, 229, 33

\bibitem[\protect\citeauthoryear{Cherepashchuk et al.}{2003}]{Cher03}
Cherepashchuk A.M., Sunyaev R.A., Seifina E.V., Panchenko I.E., Molkov S.V., 
Postnov K.A., 2003, A\&A, 411, L441

\bibitem[\protect\citeauthoryear{Cherepashchuk et al.}{2005}]{Cher05}
Cherepashchuk A.M. et al, 2005, A\&A, 437, 561

\bibitem[\protect\citeauthoryear{Cherepashchuk et al.}{2007}]{Cher06}
Cherepashchuk A.M. et al, 2007, Proc. 6th {\it INTEGRAL} Workshop, ESA SP-622, p. 319

\bibitem[\protect\citeauthoryear{Cherepashchuk et al.}{2009}]{Cher09}
Cherepashchuk A.M., Sunyaev R.A., Postnov K.A., Antokhina E.A., Molkov S.V., 
2009, MNRAS, 397, 479

\bibitem[\protect\citeauthoryear{Cherepashchuk et al.}{2012}]{Cher13}
Cherepashchuk A., Sunyaev R., Molkov S., Antokhina E., Postnov K., Bogomazov A., 
2013, in Proc. 9th {\it INTEGRAL} Workshop, 
PoS({\it INTEGRAL} 2012), id.40 (arXiv:1212.3443)

\bibitem[\protect\citeauthoryear{Collins \& Newsom}{1986}]{collins1986} Collins II G. W., Newsom G. H., ApJ, 1986, 308, 144

\bibitem[\protect\citeauthoryear{Crampton \&
Hutchings}{1981}]{crampton1981} Crampton D., Hutchings J. B., ApJ,
1981, 251, 604

\bibitem[\protect\citeauthoryear{Davydov et al.}{2008}]{Davydov08}
Davydov V.V., Esipov V.F., Cherepashchuk A.M., 2008, Astron. Rep., 52, 487

\bibitem[\protect\citeauthoryear{Fabrika}{2004}]{Fab04} Fabrika, S.N., 2004, Astrophys. Space Phys. Rev., 12, 1

\bibitem[\protect\citeauthoryear{Fabrika \& Bychkova}{1990}]{fabrika1990} Fabrika S. N., Bychkova L. V., 1990, A\&A, 240, L5

\bibitem[\protect\citeauthoryear{Filippova et al.}{2006}]{Fil06}
Filippova E., Revnivtsev M., Fabrika S., Postnov K., Seifina E., 2006, A\&A, 460, 125


\bibitem[\protect\citeauthoryear{Gehrels et al.}{2004}]{Swift}
Gehrels N. et al., 2004, ApJ, 611, 1005

\bibitem[\protect\citeauthoryear{Gies et al.}{2002a}]{gies2002} Gies D.R. 
McSwain M.V., Riddle R.L., Wang Z., Wiita P.J., Wingert D.W., 2002a, ApJ, 566,
1069 

\bibitem[\protect\citeauthoryear{Gies et al.}{2002b}]{Gies02}
Gies D.R., Huang W., McSwain M.V., 2002b, ApJ, 578, L67

\bibitem[\protect\citeauthoryear{Goranskij}{2011}]{Goransky_2011}
Goranskij V.P., 2011, Peremennye Zvezdy, 31, N5 [arXiv:1110.5304]

\bibitem[\protect\citeauthoryear{Goranskij et al.} {1998}]{Goransk98}
Goranskij V.P., Esipov V.F., Cherepashchuk A.M., 1998, Astron. Rep., 42, 209; ibid., p. 336

\bibitem[\protect\citeauthoryear{Hillwig et al.} {2004}]{Hillwig_ea04}
Hillwig T.C., Gies D.R., Huang W., 
McSwain M.V., Stark M.A., van der Meer A., Kaper L., 2004, ApJ, 615, 422

\bibitem[\protect\citeauthoryear{Hillwig \& Gies}{2008}]{hillwig08}
Hillwig T.C., Gies D.R., 2008, ApJ, 676, L37


\bibitem[\protect\citeauthoryear{Katz et al.}{1982}]{Katz_ea82}
Katz J.I., Anderson S.F., Grandi S.A., Margon B., 1982, ApJ, 260, 780

\bibitem[\protect\citeauthoryear{Kawai et al.}{1989}]{Kawai89}
Kawai N., Matsuoka M., Pan H.-C., Stewart G.C., 
1989, PASJ, 41, 491

\bibitem[\protect\citeauthoryear{Kotani}{1998}]{KotaniPhD}
Kotani T., 1998, PhD. The Institute of Space and Astronautical Science.
Japan

\bibitem[\protect\citeauthoryear{Kotani et al.}{1998}]{Kotani98}
Kotani T., Kawai N., Matsuoka M., Brinkmann W., 1998, 
in Proc. IAU Symp. 188, eds. K. Koyama, S. Kitamoto, M. Itoh,
Dordrecht: Kluwer, p. 358

\bibitem[\protect\citeauthoryear{Krivonos et al.}{2010}]{Krivonos_al10}
Krivonos R., Revnivtsev M., Tsygankov S., Sazonov S., Vikhlinin A., 
Pavlinsky M., Churazov E., Sunyaev R., 2010, A\&A, 519, A107

\bibitem[\protect\citeauthoryear{Krivosheyev et al.}{2009}]{Krivosheev09}
Krivosheyev Yu. M., Bisnovatyi-Kogan G.S., Cherepashchuk A.M., Postnov K.A., 2009, MNRAS, 394, 1674

\bibitem[\protect\citeauthoryear{Levine \& Jernigan}{1982}]{LevineJernigan82}
Levine A.M., Jernigan J.G., 1982, ApJ, 262, 294

\bibitem[\protect\citeauthoryear{Lomb}{1976}]{Lomb76}
Lomb N.R., 1976, ApSS, 39, 447

\bibitem[\protect\citeauthoryear{Margon}{1984}]{Margon84}
Margon B., 1984, ARA\&A, 22, 507

\bibitem[\protect\citeauthoryear{Margon \& Anderson}{1989}]{MargonAnderson89}
Margon B., Anderson S.F., 1989, ApJ, 347, 507

\bibitem[\protect\citeauthoryear{Molkov et al.}{2004}]{Mol04}
Molkov S., Cherepashchuk A.M., Lutovinov A.A., Revnivtsev M.G., 
Postnov K.A., Sunyaev R.A., 2004, Astron. Lett., 30, 534

\bibitem[\protect\citeauthoryear{Namiki et al.}{2003}]{Namiki03}
Namiki M., Kawai N., Kotani T., Makishima K., 2003, PASJ, 55, 281

\bibitem[\protect\citeauthoryear{Okuda et al.}{2009}]{Okuda_ea09}
Okuda T., Lipunova G.V., Molteni D., 2009, MNRAS, 398, 1668

\bibitem[\protect\citeauthoryear{Revnivtsev et al.}{2004a}]{Revnivtsev04}
Revnivtsev M. et al., 2004a, A\&A, 424, L5

\bibitem[\protect\citeauthoryear{Revnivtsev et al.}{2004b}]{Revnivtsev_al04}
Revnivtsev M.G. et al., 2004b, Astron. Let., 30, 382

\bibitem[\protect\citeauthoryear{Revnivtsev et al.}{2006}]{Revnivtsev06}
Revnivtsev M. et al., 2006, A\&A, 447, 545

\bibitem[\protect\citeauthoryear{Scargle}{1982}]{Scargle82}
Scargle J., 1982, ApJ, 263, 835

\bibitem[\protect\citeauthoryear{Shakura \& Sunyaev}{1973}]{ShS73}
Shakura N.I., Sunyaev R.A., 1973, A\&A, 24, 337

\bibitem[\protect\citeauthoryear{Trushkin et al.}{2001}]{Trushkin_ea01}
Trushkin S.A., Borisov N.N., Smirnova Yu.V., 2001, Ast. Rep., 45, 804

\bibitem[\protect\citeauthoryear{van den Heuvel et al.}{1980}]{heuvel1980} 
van den Heuvel E. P. J., Ostriker J. P., Petterson J.
A., 1980, A\&A, 1980, 81, L7

\bibitem[\protect\citeauthoryear{Whitmire \& Matese}{1980}]{whitmire1980} 
Whitmire D. P., Matese J. J., 1980, MNRAS, 193, 707




\end{thebibliography}
\end{document}